\documentclass[
a4paper,twocolumn,11pt,accepted=2026-09-23
]{quantumarticle}
\pdfoutput=1

\usepackage{dcolumn}
\usepackage[numbers]{natbib}
\usepackage{amsfonts}
\usepackage{amssymb}
\usepackage{mathrsfs}
\usepackage{mathtools}
\usepackage{amsmath}
\usepackage{amsthm}
\usepackage{graphicx}
\usepackage{color}
\usepackage{array}
\usepackage[makeroom]{cancel}
\usepackage{changes}
\usepackage{qcircuit}
\usepackage{braket}
\usepackage{bbm}
\usepackage{bm}
\usepackage{multirow}
\usepackage{cellspace} 
\usepackage{appendix}
\usepackage{tikz}
\usepackage{algorithm}
\usepackage{algpseudocode}
\usetikzlibrary{positioning}
\usetikzlibrary{shapes,arrows,arrows,positioning,fit}
\usepackage{orcidlink}
\usepackage{comment}
\usepackage[export]{adjustbox}

\usepackage[capitalize]{cleveref}
\crefname{section}{Sec.}{Secs.}

\crefname{thm}{Theorem}{Theorems}
\crefname{dfn}{Definition}{Definitions}
\crefname{rmk}{Remark}{Remarks}
\crefname{lem}{Lemma}{Lemmas}
\crefname{cor}{Corollary}{Corollaries}

\theoremstyle{plain}

\theoremstyle{remark}

\begin{document}

\title{Fast simulations of X-ray absorption spectroscopy for battery materials on a quantum computer}

\author{Stepan Fomichev \orcidlink{0000-0002-1622-9382}}
\thanks{These authors contributed equally.\\
stepan.fomichev@xanadu.ai\\
pablo.casares@xanadu.ai}
\affiliation{Xanadu, Toronto, ON, M5G 2C8, Canada}

\author{Pablo A. M. Casares \orcidlink{0000-0001-5500-9115}}
\thanks{These authors contributed equally.\\
stepan.fomichev@xanadu.ai\\
pablo.casares@xanadu.ai}
\affiliation{Xanadu, Toronto, ON, M5G 2C8, Canada}

\author{Jay Soni}
\affiliation{Xanadu, Toronto, ON, M5G 2C8, Canada}

\author{Utkarsh Azad \orcidlink{0000-0001-7020-0305}}
\affiliation{Xanadu, Toronto, ON, M5G 2C8, Canada}

\author{Alexander Kunitsa \orcidlink{0000-0002-3640-8548}}
\affiliation{Xanadu, Toronto, ON, M5G 2C8, Canada}

\author{Arne-Christian Voigt}
\affiliation{Volkswagen AG, Berliner Ring 2, 38440 Wolfsburg, Germany} 

\author{Jonathan E. Mueller }
\affiliation{Volkswagen AG, Berliner Ring 2, 38440 Wolfsburg, Germany} 

\author{Juan Miguel Arrazola \orcidlink{0000-0002-0619-9650}}
\affiliation{Xanadu, Toronto, ON, M5G 2C8, Canada}

\begin{abstract}

X-ray absorption spectroscopy (XAS) is a leading technique for understanding structural changes in advanced battery materials such as lithium-excess cathodes. However, extracting critical information like oxidation states from the experimental spectra requires expensive and time-consuming simulations. 
Building upon a recent proposal to simulate XAS using quantum computers, this work proposes a highly-optimized implementation of the time-domain algorithm for X-ray absorption.
Among a host of improvements to Hamiltonian representation, circuit implementation, and measurement strategies, three optimizations are key to the efficiency of the algorithm. The first is the use of product formulas with the compressed double factorized form of the Hamiltonian. 
The second is recognizing that for spectroscopy applications, it is sufficient to control the error in the eigenvalues of the (approximate) Hamiltonian being implemented by the product formula, rather than the generic error on the full time evolution operator. Using perturbation theory to estimate this eigenvalue error, we find that significantly fewer Trotter steps are needed than expected from the time evolution error bound.
The third is the choice of an optimized distribution of samples that takes advantage of the exponentially decaying Lorentzian kernel.
Through constant factor resource estimates, we show that a challenging model Li$_4$Mn$_2$O cluster system with 18 spatial orbitals and 22 electrons in the active space can be simulated with 100 logical qubits and less than $4 \times 10^8$ Toffoli gates per circuit. 
Finally, the algorithm is implemented on a simulator, and the reconstructed spectrum is verified against a classical computational reference. 
The low cost of our algorithm makes it attractive to use on fault-tolerant quantum devices to accelerate the development and commercialization of high-capacity battery cathodes.

\end{abstract}

\maketitle

\section{Introduction}

Reliable high-capacity energy storage is a crucial technology, especially in light of on-going efforts to increase the adoption of electric vehicles. A promising approach to increasing battery capacity is the use of lithium-excess materials for the battery cathode. Li-excess materials are transition metal oxides whose crystal lattices have been engineered to accommodate a larger lithium content, typically through selective substitution in the transition metal layers \cite{lu2002understanding,lu2002synthesis,saubanere2016intriguing,tran2008mechanisms,zhang2022pushing,hong2015lithium,hy2016performance,radin2017narrowing}. Unfortunately, most prototype Li-excess cathodes to date have tended to undergo rapid performance decline with each charge and discharge cycle, linked to irreversible structural degradation of the crystal structure~\cite{johnson2008synthesis,bettge2013voltage}. A deep understanding of the degradation processes, specifically of the local bonding structure in the bulk of these materials, could enable scientists to address this challenge and produce reliable high-capacity cathodes~\cite{fomichev2024simulating}. 

There are several competing hypotheses for the degradation mechanism. One explanation is that the crystal lattice degrades through oxygen dimerization: when Li is depleted from the cathode beyond a certain point, the remaining loose bonds of crystal oxygens re-align, resulting in formation of oxygen peroxides or even molecular oxygen \cite{seo2016structural,luo2016charge,koga2013different,koga2014operando,house2023delocalized}. Alternative points of view instead attribute the degradation to changing oxidation states of the transition metals \cite{kalyani1999lithium,ohzuku2011high,radin2019manganese}. 

Local electronic structure of bulk materials is notoriously difficult to probe, especially during the charging-discharging cycles: yet that is precisely the information that would shed light on structural degradation of Li-excess cathodes over the course of several charging cycles  \cite{radin2019manganese,house2023delocalized}. X-ray absorption spectroscopy (XAS), specifically X-ray absorption near-edge spectroscopy (XANES), is a leading method for studying local electronic structure. This is due to its ability to penetrate sample bulk, its high elemental specificity, and its direct connection to local structure through the excitation of tightly bound, localized core electrons~\cite{de2001high}. However, teasing out mechanism-relevant quantities such as oxidation states from the spectra is highly nontrivial without reliable complementary simulations, for which existing methods either lack the necessary level of accuracy (such as those based on density functional theory \cite{stener2003time,ray2007description,besley2010time,liang2011energy,lopata2012linear,lestrange2015calibration,stetina2019modeling}) or are simply prohibitively expensive to run for the needed system sizes (such as wavefunction-based methods \cite{roos1980complete,casanova2009restricted,maganas2014combined,yang2014multireference,pinjari2014restricted,guo2016simulations,sassi2017first,delcey2019efficient,maganas2019comparison,guo2020restricted,lee2023ab,casanova2022restricted}). This is particularly the case when strongly-correlated transition metals are present, as is typically the case with lithium-excess materials. There is thus a strong need for high-accuracy and performant simulation methods to resolve outstanding questions regarding structural degradation of Li-excess cathodes -- a challenge that quantum computers could be well-poised to solve \cite{fomichev2024simulating}. 

\begin{figure*}
    \centering
    \includegraphics[width=0.4\textwidth]{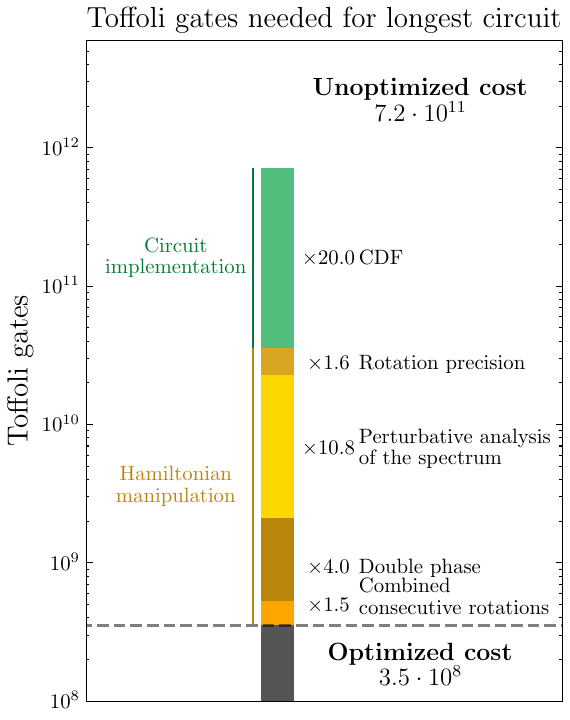}
    \includegraphics[width=0.4\textwidth]{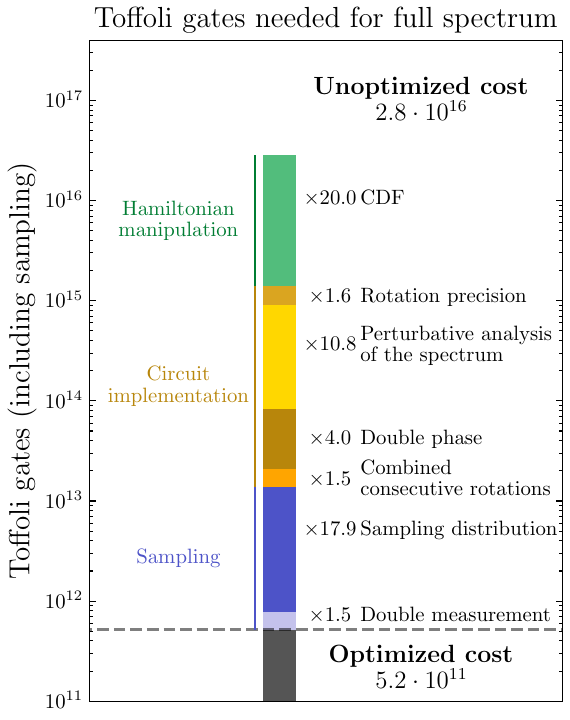}
    \caption{\textbf{Left: Toffoli gates in longest circuit.} Number of Toffoli gates required to run the longest circuit for the time-domain XAS algorithm for the CAS(22e, 18o) Li$_4$Mn$_2$O cluster, as indicated in~\cref{tab:resources}. We depict the impact of different circuit optimizations. \textbf{Right: Toffoli gates for full spectrum with sampling.} Number of Toffoli gates required to run the time-domain XAS algorithm for the CAS(22e, 18o) Li$_4$Mn$_2$O cluster, as indicated in~\cref{tab:resources}. We depict the impact of different circuit optimizations discussed in the paper.}
    \label{fig:optimizations_and_cost}
\end{figure*}

In this paper, we introduce a highly-optimized quantum algorithm for simulating X-ray absorption spectra, based on the time-domain approach of Fomichev \textit{et al.} ~\cite{fomichev2024simulating}. This algorithm makes quantum-based simulation of XAS spectra an enticing prospect on upcoming generations of early fault-tolerant hardware.
To demonstrate the efficiency of the proposed algorithm, we provide a constant-factor resource estimation for the Li$_4$Mn$_2$O cluster active space CAS(22e, 18o). This system was identified in Ref.~\cite{fomichev2024simulating} as a prototypical example of a problem currently beyond the reach of existing classical simulation methods, yet directly industrially relevant to the challenge of structural degradation in Li-excess as a representative structure that  arises in delithiation events.

To achieve low cost, we introduce and exploit several key optimizations to the algorithm. The first is the use of product formulas with the compressed double factorized form of the Hamiltonian for time evolution: this reduces the number of terms to be implemented with the product formula relative to the direct Jordan-Wigner approach. The second is the method for determining the number of product formula steps required for the spectroscopy application. In the case of spectroscopy, controlling the generic error in implementing the full time evolution operator is unnecessarily stringent: instead, it is sufficient to make sure that the eigenvalues of the Hamiltonian being implemented through product formulas are within the error threshold of the true eigenvalues. Estimating the eigenvalue deviation by using perturbation theory, we are able to dramatically reduce the number of Trotter steps expected to be necessary for the application. The third optimization is employing a sampling distribution that takes advantage of the decaying Lorentzian kernel to exponentially reduce shot allocation to longer evolution times. When combining these and numerous other improvements of the time evolution circuit implementation, measurement schemes, and Hamiltonian representation, we demonstrate final Toffoli-gate counts as low as $3.54 \times 10^{8}$ for the longest circuit, and only 100 workspace logical qubits. Altogether the optimizations achieve five orders of magnitude gate count reduction -- their relative contributions are shown in \cref{fig:optimizations_and_cost} and computed in \cref{sec:application}. When implemented in an active volume architecture \cite{litinski2022active} with a total of 350 logical qubits, we require around $1.08 \times 10^{8}$ cycles for the longest circuit, and $1.58 \times 10^{11}$ total logical cycles, meaning a runtime of less than two days on a quantum computer operating at MHz logical clock rates.

The manuscript is organized as follows. In \cref{sec:xas}, we outline the simulation problem of obtaining the XAS absorption cross-section. We then proceed in \cref{sec:optimizations} to thoroughly describe the quantum algorithm together with the key optimizations that allowed us to achieve low implementation cost. 
In \cref{sec:resources}, we perform constant-factor resource estimate analysis of the algorithm and use error analysis to constrain the free parameters like maximal evolution time and the sampling budget required.
With the algorithm and resource estimate expressions in hand, in \cref{sec:application} we apply the algorithm to the prototypical Li$_4$Mn$_2$O CAS(22e,18o) model cluster: we benchmark the algorithm performance by implementing it on a simulator, and present constant factor resource estimates for this system, highlighting the effect of each specific improvement on the overall cost. Finally, in \cref{sec:conclusions}, we present our conclusions. A tutorial to replicate the implementation of the algorithm here might be found in~\url{https://pennylane.ai/qml/demos/tutorial_xas}.

\section{Algorithm for X-ray absorption\label{sec:xas}}

\subsection{Simulation problem}

In XAS experiments, the quantity of interest is the absorption cross section $\sigma_A(\omega)$. We focus on the absorption $K$-edge — that is, the one driven by excitation of the core $1s$ electrons. Further, we will analyze the range of X-ray energies where the dipole approximation remains valid, and neglect any relativistic corrections (although they can be added straightforwardly in our formalism). 
Using Fermi’s golden rule, it can be shown that the absorption cross-section is given by
\begin{equation}
    \sigma_A(\omega) = \frac{4\pi}{3\hbar c}\omega \sum_{F \neq I} \sum_{\rho = x, y, z}\frac{| \bra{F}\hat{m}_{\rho}\ket{I} |^2 \eta}{(E_F - E_I - \omega)^2 + \eta^2}.
    \label{eq:crosssection}
\end{equation}
Here $\ket{I}$ and $\ket{F}$ are respectively the many-body ground and excited states (with $E_I$ and $E_F$ their respective energies) of the system electronic Hamiltonian $\hat H$, $\omega$ is the angular frequency of the incoming X-ray, $\eta$ is the line broadening which we set by the experimental resolution (often around 1 eV), and $\hat{m}_\rho$ is the $\rho$-th Cartesian component of the dipole operator for the molecular cluster. The electronic Hamiltonian is given by
\begin{multline}\label{eq:el-ham}
    \hat{H} = E + \sum_{p,q = 1}^N \sum_{\gamma \in \{\uparrow,\downarrow\}} (p|\kappa|q) a_{p\gamma}^\dagger a_{q\gamma}+ \\ \frac{1}{2}\sum_{p,q,r,s=1}^N\sum_{\gamma,\tau \in \{\uparrow,\downarrow\}}(pq|rs) a_{p\gamma}^\dagger a_{q\gamma} a_{r\tau}^\dagger a_{s\tau},
\end{multline}
where $a_{p\gamma}^{(\dagger)}$ is the annihilation (creation) operator for spatial orbital $p$ and spin $\gamma$, $E$ is the energy offset, $N$ is the number of spatial orbitals, and $(p|\kappa|q)$ and $(pq|rs)$ are the one- and two-electron integrals, respectively, given by ~\cite{cohn2021quantum}
\begin{align}
    (pq|rs) &\equiv \iint_{\mathbb{R}^6} d\mathbf{r}_1 d\mathbf{r}_2\, \phi_p(\mathbf{r}_1) \phi_q(\mathbf{r}_1) \nonumber\\
    &\qquad\times \frac{1}{r_{12}} \phi_r(\mathbf{r}_2) \phi_s(\mathbf{r}_2).\\
    (p|\kappa|q) &\equiv (p|-\frac{1}{2}\nabla_1^2|q) + (p|-\sum_A \frac{Z_A}{r_{1A}}|q)\\
    &+ 2 \sum_i (pq|ii) - \frac{3}{2} \sum_r (pr|qr),
\end{align}
with $\phi_p(\mathbf{r})$ being the $p$'th spatial molecular orbital, $Z_A$ being the atomic number of the $A$'th atom, and $r_{12} = |\mathbf{r}_1 - \mathbf{r}_2|, r_{1A} = |\mathbf{r}_1 - \mathbf{r}_A|$ the distances between electronic positions $\mathbf{r}_i$ and nuclei positions $\mathbf{R}_A$. 

\begin{figure}[t]
    \centering
    \includegraphics[width=0.6\linewidth]{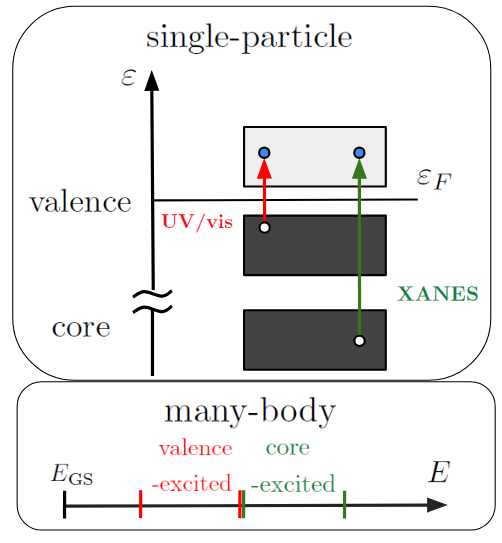}
    \caption{Schematic of electronic excitations relevant in XAS absorption. High-energy X-ray photons result in the creation of core holes, creating many-body excited states that are energetically high above the numerous valence-excited states probed in optical absorption experiments.}
    \label{fig:core-excited-problem}
\end{figure}

What distinguishes XAS from other absorption spectroscopies governed by a similar expression is the nature of the excited states $\ket{E_F}$ relevant in the spectrum. Given that X-rays excite core electrons, the main contributing excited states will be the core-excited states — highly excited states that lie significantly above the low-lying valence-excited states in energy (see \cref{fig:core-excited-problem}). Such states typically exhibit strong correlations, especially in a system with transition metals. In this study, we use the core-valence separation approximation \cite{cederbaum1980many,barth1981many,norman2018simulating,herbst2020quantifying} to allow us to separate the core-excited and valence-excited state manifolds and avoid spending resources on computing peaks in the valence-excited sector of the spectrum: we will elaborate on the practical implementation of the scheme in \cref{sec:application}.

The goal of obtaining the absorption cross-section numerically is to use it to interpret experimental XAS measurements. All the degradation mechanisms mentioned in the introduction are associated with specific oxidation states of oxygen and the transition metal: using XAS to deduce those oxidation states can thus help infer the degradation mechanism. Simulations allow us to study absorption cross-sections for various possible oxidation states, by modifying the local coordination environment of the absorbing atom. By finding the simulated spectrum that best matches experiment -- i.e. performing spectral fingerprinting -- we can determine the oxidation state, and thus gain information regarding the mechanism of structural degradation. 

\subsection{Time-domain algorithm}

\begin{figure}[t]
    \centering
    \[
    \Qcircuit @C=1em @R=1em {
    \lstick{\ket{0}} & \gate{H} & \gate{S^\dagger} & \ctrl{1} & \gate{H} & \meter \gategroup{1}{3}{1}{3}{.7em}{--} \\
    \lstick{\frac{\hat{m}_\rho\ket{I}}{\|\hat{m}_\rho\ket{I}\|}} & {/}\qw & \qw & \gate{e^{-i\tau j H}} & \qw & \qw \\
    }
    \]
    \caption{Circuit evaluating the time evolution expectation values that encode the XAS absorption spectrum. Upon loading the initial state $\hat{m}_\rho\ket{I}/\|\hat{m}_\rho\ket{I}\|$, we perform the Hadamard test on the unitary for time evolution under the system Hamiltonian. With the phase gate $S^{\dagger}$ present (absent), this gives the real (imaginary) part of the Green's function $\tilde{G}(\tau j)$ (\cref{eq:G(t)}). The evolution times $j$ are sampled via a Monte Carlo procedure: see the text for details.}
    \label{fig:algo-circs-timedomain}
\end{figure}
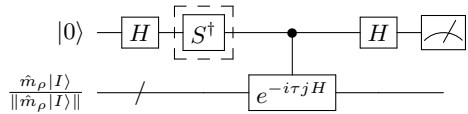

The cross-section in~\cref{eq:crosssection} can be viewed as the imaginary part of the frequency-domain Green’s function
\begin{equation}
    \mathcal{G}_\rho(\omega) = \bra{I} \hat{m}_{\rho} \frac{1}{\hat{H} - E_I 
    - \omega + i\eta} \hat{m}_{\rho} \ket{I},
    \label{eq:g-freq-unnormalized}
\end{equation}
up to multiplicative prefactors. This can be seen by inserting the resolution of identity to the right of the resolvent and taking the imaginary part of the result. The first step gives 
\begin{multline}
    \bra{I} \hat{m}_{\rho} \frac{1}{\hat{H} - E_I 
    - \omega + i\eta} \left(\sum_F \ket{F} \bra{F} \right) \hat{m}_{\rho} \ket{I} = \\
    = \sum_F \frac{|\bra{F} \hat{m}_\rho \ket{I}|^2}{(E_F - E_I - \omega + i\eta)}.
\end{multline}
Now taking the imaginary part and using the relation 
\begin{equation}
    \text{Im}\left(\frac{1}{x+ i \eta} \right) = - \frac{\eta}{x^2+\eta^2},
\end{equation}
we arrive at
\begin{multline}\label{eq:cross_section_green}
    \operatorname{Im}\mathcal{G}_\rho(\omega) = -\sum_{F \neq I} \frac{|\bra{F} \hat{m}_\rho \ket{I}|^2 \eta}{(E_F - E_I - \omega)^2 + \eta^2} \\
    - \frac{|\bra{I}\hat{m}_\rho\ket{I}|^2 \, \eta}{\omega^2 + \eta^2}.
\end{multline}
Comparing \cref{eq:cross_section_green} and \cref{eq:crosssection}, we recognize that we have recovered the absorption cross-section up to multiplicative constants and an additive term that can be removed in pre-processing by subtracting from the dipole operator its expectation value in the initial state, $\hat{m}_\rho \rightarrow \hat{m}_\rho - \bra{I} \hat{m}_\rho \ket{I}$.
Since quantities evaluated in the quantum register have to be normalized, we also define the normalized Green's function $G_\rho(\omega)$ as 
\begin{equation}
G_\rho(\omega) = \frac{\mathcal{G}_\rho(\omega)}{\|\hat{m}_\rho \ket{I} \|^2}.
\end{equation}
The factor of $1/\|\hat{m}_\rho \ket{I} \|^2$ make sure that $\eta \ |G_\rho(\omega)|\leq 1$.
Instead of evaluating this Green's function directly in frequency space, we follow the time domain approach, as time evolution is a more natural quantum computing operation than matrix inversion and thus often cheaper to perform \cite{fomichev2024simulating}. The imaginary part of the frequency domain Green’s function may be written as a discrete-time Fourier transform
\begin{equation}
    - \eta \operatorname{Im} G_\rho(\omega) = \frac{\eta\tau}{2\pi} \sum_{j=-\infty}^{\infty} e^{-\eta\tau|j|} \tilde{G}(\tau j) e^{ij\tau\omega},
    \label{eq:green-via-discretetime-fourier}
\end{equation}
where $\tau \sim O(\|\hat H\|^{-1})$ is chosen such that all the eigenstates $\ket{F}$ on which the state $\hat{m}_{\rho} \ket{I}$ has support are re-scaled into the range $[-\pi, \pi)$, and we periodized the Lorentzian as $1/(x^2 + \eta^2) \rightarrow \sum_j 1 / ((x-2\pi j)^2+\eta^2)$ (following the derivation in Ref. \cite{fomichev2024simulating}). 
Here $\tilde{G}_\rho(\tau j)$ is the discrete-time time-domain (normalized) Green's function, given by 

\begin{equation}\label{eq:G(t)}
    \tilde{G}_\rho(\tau j) = \frac{\bra{I} \hat{m}_\rho e^{-i \hat H \tau j} \hat m_{\rho} \ket{I}}{\|\hat m_\rho \ket{I} \|^2}.
\end{equation}
This expectation value can be evaluated directly for varying times $j$ using the Hadamard test circuit (\cref{fig:algo-circs-timedomain}). The initial state $\hat m_\rho \ket{I}$ can be obtained by using classical methods to find the ground state $\ket{I}$ and then applying the dipole operator to it, with the resultant state prepared in the register using the sum-of-Slaters method from Ref.~\cite{fomichev2024initial}. Fourier-transforming $\tilde{G}(\tau j)$ classically yields $G(\omega)$ and thus $\sigma_A(\omega)$. 

\section{Algorithm implementation\label{sec:optimizations}}

\subsection{Time evolution via product formulas with compressed double factorization\label{ssec:cdf}}

The key subroutine needed to compute the matrix elements in \cref{eq:G(t)} is time evolution $e^{-i \hat{H}  \tau j}$ under the electronic Hamiltonian. In this work we focus on the Trotter product formula approach to time evolution, which offers a reduced logical qubit overhead compared to qubitization methods. Implementing time evolution using Trotter product formulas requires decomposing the Hamiltonian into a sum of efficiently simulable (fast-forwardable) terms, or fragments. Instead of directly using the Jordan-Wigner mapping to generate such a decomposition, we employ the compressed double factorization (CDF) technique (also called the Cartan subalgebra approach)~\cite{yen2021cartan,cohn2021quantum,oumarou2024accelerating}. 
As we quantify more explicitly through constant-factor resource estimation in \cref{sec:resources}, the CDF approach results in the total number of fragments scaling as $\tilde{O}(N^3)$, rather than $O(N^4)$ obtained with direct Jordan-Wigner mapping, and overall gives a low-cost implementation of time evolution based on product formulas.

The main idea behind CDF is to decompose the Hamiltonian into a collection of fragments, and to find a single-particle basis for each fragment in which it is diagonal (and thus simple to fast-forward). Time evolution under each fragment can be performed by first rotating to the fragment's diagonalizing basis, implementing that fragment, then rotating back. 

\textbf{One-body fragment:} To find the single-particle basis that diagonalizes the one-body part of the electronic Hamiltonian in \cref{eq:el-ham}, we diagonalize the one-electron integral matrix 
\begin{equation}
    (p|\kappa|q) = \sum_k \tilde{U}_{pk}^{(0)} \tilde{Z}_{kk}^{(0)} \tilde{U}_{qk}^{(0)}.
\end{equation}
This gives a matrix of eigenvalues $\tilde{Z}_{kk}^{(0)}$, and an orthogonal matrix of eigenvectors $U_{pq}^{(0)}$ that transforms the creation and annihilation operators to the new basis, which we label with the superscript $(0)$,
\begin{equation}
    a^{(0)\dagger}_{k\gamma} = \sum_{p} U_{pk}^{(0)} a^\dagger_{p\gamma}, \quad a^{(0)}_{k\gamma} = \sum_{q} U_{qk}^{(0)} a_{q\gamma}.
\end{equation}
Using these orthogonal matrices, we rewrite the one-electron Hamiltonian term as
\begin{align}
    \sum_{p,q = 1}^N &\sum_{\gamma \in \{\uparrow,\downarrow\}} (p|\kappa|q) a_{p\gamma}^\dagger a_{q\gamma}\nonumber = \\
    =& \sum_{p,q=1}^N \sum_{\gamma\in\{\uparrow,\downarrow\}} \sum_{k=1}^N\tilde{U}_{pk}^{(0)} \tilde{Z}_{kk}^{(0)} \tilde{U}_{qk}^{(0)} \nonumber\\
    &\times a^\dagger_{p\gamma} a_{q\gamma}\nonumber = \\
    =&\sum_{k=1}^N \sum_{\gamma\in\{\uparrow,\downarrow\}} \tilde{Z}_{kk}^{(0)} \left(\sum_{p=1}^N \tilde{U}_{pk}^{(0)} a^\dagger_{p\gamma} \right) \left(\sum_{q=1}^N \tilde{U}_{qk}^{(0)} a_{q\gamma}\right)\nonumber 
    \\
    &=\sum_{k=1}^N \sum_{\gamma\in\{\uparrow,\downarrow\}} \tilde{Z}_{kk}^{(0)} a_{k\gamma}^{(0)\dagger} a^{(0)}_{k\gamma}. 
    \label{eq:onebody-factorized}
\end{align}

Implementing the time evolution under this operator requires two steps. First, we rotate between the $(0)-$basis (where the operator is diagonal) and the computational basis in which the initial state is expressed.
The unitary $\bm{U}^{(0)}$ that will transform the initial many-qubit state in response to rotating the single-particle basis by $U_{pq}^{(0)}$ can be constructed using Thouless's theorem \cite{kivlichan2018quantum,thouless1960stability},
\begin{align}
    \bm{\tilde{U}}^{(0)} = \exp\left(\sum_{p,q}[\log \tilde{U}^{(0)}]_{pq} (a^\dagger_p a_q -a^\dagger_q a_p) \right),
\end{align}
and subsequently decomposing it into Givens rotations \cite{arrazola2022universal}. 
Once in the $(0)-$basis, we apply the Jordan-Wigner transformation to map the resulting particle number operator $a_{k,\gamma}^\dagger a_{k\gamma} = n_{k\gamma}$ to Pauli operators, via $n_k = (1-\sigma_{z,k\gamma})/2$, where $\sigma_{z,k\gamma}$ is the Pauli $Z$ operator for spin-orbital $(k,\gamma)$. As such, the time evolution of~\cref{eq:onebody-factorized} can be implemented as
\begin{multline}\label{eq:one_body_fragment}
    \exp\left({-i \sum_{p,q,\gamma} (p|\kappa|q) a_{p\gamma}^\dagger a_{q\gamma}}\right)\\
    = \bm{\tilde{U}}^{(0)} \left[\prod_{k,\gamma} \exp\left(-i \frac{\tilde{Z}_{kk}^{(0)}}{2} \sigma_{z,k\tau}\right)\, \right]\left(\bm{\tilde{U}}^{(0)}\right)^T,
\end{multline}
where we have factored out the global phase factor of $\exp({-i\sum_k \tilde{Z}_{kk}^{(0)} / 2})$.
See the first part of the circuit in \cref{fig:CDF} for a visual illustration of the implementation. The implementation of $\bm{\tilde{U}}^{(0)}$ can be carried out with $O(N^2)$ Givens rotations, at depth $O(N)$~\cite{kivlichan2018quantum,arrazola2022universal}. We also need $N$ single-qubit Pauli $Z$ rotations to implement $\left[\prod_{k,\gamma} \exp(-i \tilde{Z}_{kk}^{(0)} \sigma_{z,k\tau}/2)\, \right]$.
Note that since the electronic Hamiltonian preserves spin, the unitary can be directly built to be block-diagonal over the two spin sectors.

\textbf{Two-body fragments:} Rather than trying to directly diagonalize the two-electron integral tensor $(pq|rs)$, as we did with the one-electron integral term, we ``factorize'' it approximately: adopting the CDF factorization ansatz, we write
\begin{align}\label{eq:(pq|rs)}
   (pq|rs) \approx \sum_{\ell=1}^L \sum_{k,l=1}^N U^{(\ell)}_{pk} U^{(\ell)}_{qk} Z^{(\ell)}_{kl} U^{(\ell)}_{rl} U^{(\ell)}_{sl}.
\end{align}
Here $Z^{(\ell)}_{kl}$ are the entries of symmetric matrices $Z^{(\ell)}$ for each $\ell$, and $U^{(\ell)}_{pq}$ are entries of orthogonal matrices~\cite{cohn2021quantum}; this contraint is enforced in practice by defining $U^{(\ell)}$ through the matrix exponential of an antisymmetric matrix, $U^{(\ell)} = \exp(X^{(\ell)})$ with $(X^{(\ell)})^T = -X^{(\ell)}$. The matrices $X^{(\ell)}$ and $U^{(\ell)}$ are unknown \textit{a priori}, and are determined by solving the optimization problem of minimizing the difference between the left and right hand sides of \cref{eq:(pq|rs)}~\cite{oumarou2024accelerating}. In practice, we find that the optimization procedure is fairly robust to the choice of the optimization schedule, and a random initialization of the matrices works well. More details on the implementation of the CDF optimization, as well as a PennyLane \cite{bergholm2018pennylane} code tutorial, are available in Ref. \cite{UtkarshAzad2025}.

The number of fragments $L$ determines the accuracy of the decomposition. Previous studies have found that fixing $L = \tilde{O}(N)$ is often sufficient to achieve constant and reasonable error with increased system size~\cite{motta2021low,von2021quantum,lee2021even}, which we confirmed for the Li$_4$Mn$_2$O cluster, see \cref{app:CDF} and specifically \cref{fig:CDF-error-LiMnO}. In practice, we found it beneficial to perform optimization of the $X^{(\ell)}$ and $Z^{(\ell)}$ matrices fragment by fragment: sequentially adding and optimizing new fragments while keeping prior ones frozen --- higher values of $\ell$ end up corresponding to smaller corrections of the CDF Hamiltonian.  

 Once the single-particle basis rotations $U^{(\ell)}$ and the $Z^{(\ell)}$ matrices are determined, implementing time evolution of a single two-body fragment is analogous to the one-body case. Details of the derivation of the qubit-mapped CDF Hamiltonian in terms of the $U^{(\ell)}$ and $Z^{(\ell)}$ matrices can be found in~\cref{app:CDF}. The final form of the two-body part of the Hamiltonian is
 \begin{multline}
     \sum_{p,q,r,s =1}^N\sum_{\gamma,\tau \in \{\uparrow,\downarrow\}}(pq|rs) a_{p\gamma}^\dagger a_{q\gamma} a_{r,\tau}^\dagger a_{s\tau} \\ 
     \approx \frac{1}{8}\sum_\ell  \sum_{(k,\gamma)\neq(l,\tau)}  Z^{(\ell)}_{kl} \sigma_{z,k\gamma}^{(\ell)} \sigma_{z,l\tau}^{(\ell)} \\
     - \frac{1}{2}\sum_\ell \sum_{k,l,\gamma}  Z^{(\ell)}_{kl} \sigma_{z,k\gamma}^{(\ell)}.
\label{eq:two_body_term_maintext}
 \end{multline}
 where we once again factored out a global phase factor for clarity of presentation (see \cref{app:CDF} for the full expression of the phase). As before, the $\sigma_{z,k\gamma}^{(\ell)}$ are Pauli $Z$ operators in the $(\ell)$-basis: to implement time evolution with respect to these fragments, we again rely on the unitaries $\bm{U}^{(\ell)}$ obtained via the Thouless theorem. The two-$Z$ term in the two-body fragment is implemented as
 \begin{multline}
    \exp\left(-i\sum_{(k,\gamma)\neq(l,\tau)}\frac{Z^{(\ell)}_{kl}}{8} \sigma_{z,k\gamma}^{(\ell)} \sigma_{z,l\tau}^{(\ell)}\right)\\
     = \bm{U}^{(\ell)} \left[\prod_{\substack{k,l\\\gamma, \tau}} \exp\left(-i \frac{Z_{kl}^{(\ell)}}{8} \sigma_{z,k\gamma} \sigma_{z,l\tau}\right) \right]\left(\bm{U}^{(\ell)}\right)^T.
     \label{eq:two_body_fragment}
 \end{multline}
 As above, the $\bm{U}^{(\ell)}$ are implemented via Givens rotations~\cite{kivlichan2018quantum,arrazola2022universal}. See the right side of \cref{fig:CDF} for a visual depiction. The two-qubit Pauli rotations $\sigma_{z,k\gamma}  \sigma_{z,l\tau}$ can be simplified by mapping them into a single-qubit rotation via CNOTS, see \cref{fig:rzz}.
 When it comes to the one-$Z$ term, since any one-body operator is exactly diagonalizable, in practice we group all one-$Z$ terms arising from the two-electron integral term together with those of the one-electron, and diagonalize the resulting operator at once, obtaining $U^{(0)}, Z^{(0)}$ (rather than the $\tilde{U}^{(0)}, \tilde{Z}^{(0)}$ matrices we used above for diagonalizing only $(p|\kappa|q)$). The resulting one-body operator is simulated as indicated in~\cref{eq:one_body_fragment}.
The implementation of individual fragments in \cref{eq:one_body_fragment} and \cref{eq:two_body_fragment} can then be used to perform any product formula: in this work we primarily employ the second-order Trotter product formula.
 
 \begin{figure}[t]
    \centering
    \begin{minipage}{0.45\columnwidth}
    \Qcircuit @C=0.6em @R=.7em {
     & \gate{(\bm{U}^{(0)})^T} & \multigate{1}{\prod\limits_{k\gamma} e^{i Z^{(0)}_{kk}\sigma_{z,k\gamma}}}  & \gate{\bm{U}^{(0)}} & \qw \\
   & \gate{(\bm{U}^{(0)})^T} & \ghost{\prod\limits_{k\gamma} e^{i Z^{(0)}_{kk}\sigma{z,k\gamma}}} &\gate{\bm{U}^{(0)}} & \qw
    }
    \end{minipage} \\[0.5cm]
    \begin{minipage}{0.45\columnwidth}
    \Qcircuit @C=0.6em @R=.7em {
     &  \gate{(\bm{U}^{(\ell)})^T } & \multigate{1}{\prod\limits_{(k,\gamma)\neq (l,\tau)} e^{i Z^{(\ell)}_{kl} \sigma_{z,k\gamma} \sigma_{z,l\tau}}} & \gate{\bm{U}^{(\ell)}} & \qw & \text{...}  \\
   &   \gate{(\bm{U}^{(\ell)})^T } & \ghost{\prod\limits_{(k,\gamma)\neq (l,\tau)} e^{i Z^{(\ell)}_{kl} \sigma_{z,k\gamma}\,\, \sigma_{z,l\tau}}} & \gate{\bm{U}^{(\ell)}} & \qw & \text{...}
    }
   \end{minipage}
    \caption{Circuits implementing the one-body fragments (top) and two-body fragments (bottom) arising in the compressed double factorization form of the electronic Hamiltonian~\cite{cohn2021quantum}. The register is split into two wire groups corresponding to the spin up and spin down sectors. Here $Z^{(\ell)}$ are symmetric matrices, $\sigma_{z,k\gamma}$ is the Pauli $Z$ operator acting on the $(k,\gamma)$-th spin orbital, and $\bm{U}^{(\ell)}$ are the unitaries transforming the system register according to the effect of the corresponding single-particle basis rotation $U^{(\ell)}$ in~\cref{eq:(pq|rs)}. 
    }
    \label{fig:CDF}
\end{figure}
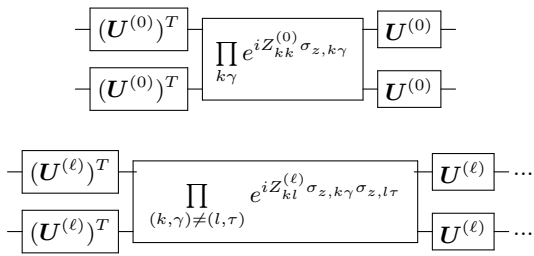

\begin{figure}[t]
    \centering
    \Qcircuit @C=1em @R=.7em {
     & \qw  & \multigate{1}{e^{-i \frac{\theta}{2}  z_k z_l}}  & \qw & &  =  &  & & \qw & \ctrl{1} & \qw & \ctrl{1} &\qw &\\
     & \qw  & \ghost{e^{-i \frac{\theta}{2}  z_k z_l}}  & \qw & &  & & & \qw & \targ & \gate{R_z(\theta)} & \targ &\qw &
}
    \caption{Mapping a two-qubit $\sigma_{z,k} \sigma_{z,l}$ rotation into a single qubit rotation via CNOTs, see~\cite[Fig. 2]{anastasiou2024tetris} for a generalization to arbitrary Pauli string rotations.}
    \label{fig:rzz}
\end{figure}
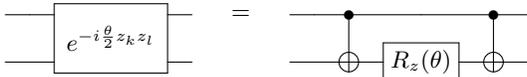

Having described the implementation of time evolution using a CDF-based product formula approach, we finish this section by describing five additional improvements to this general strategy:
\begin{enumerate}
\item \textbf{Choice of product formula:} There is a variety of product formulas available for Hamiltonian simulation: the standard Trotter-Suzuki hierarchy~\cite{suzuki1992general}, multi-product formulas and extrapolation~\cite{low2019well,blanes2024generalized,endo2019mitigating,vazquez2022enhancing,vazquez2023well,zhuk2023dynamicMPFs,robertson2024dynamicMPFs,rendon2024improved}, composition of product formulas with qDRIFT~\cite{campbell2019random,hagan2023composite,pocrnic2024composite}, and product formulas specifically designed for particular Hamiltonian structure (interaction picture, near-integrability)~\cite{blanes2000processing,blanes2004numerical,sharma2024hamiltonian,bosse2024efficient}. We explored many of these possibilities.~\cref{app:qDRIFT_&_PFs}, for instance, presents an analysis on how to compose qDRIFT with arbitrary product formulas.
However, for the relatively loose target error requirements of 1 eV driven by typical line broadening in XAS, we found empirically that a randomized second-order Trotter formula provides an optimal balance of precision and cost~\cite{strang1968construction,childs2019random}. 

\begin{figure}[t]
    \centering
    \begin{minipage}{0.7\columnwidth}
    \Qcircuit @C=0.8em @R=.7em {
     & \qw  & \ctrl{2} & \qw & \qw & & & & & \qw & \ctrl{2} & \qw & \ctrl{2} &\qw &\\
     & & & & & & \Rightarrow &&&&&\\
     & \qw  & \gate{R_z(2\theta)} & \qw & \qw & &  & & & \qw & \targ & \gate{R_z(\theta)} & \targ &\qw &\\
}    \end{minipage}

    \caption{Double phase trick: we can use cheap CNOT gates to avoid the cost of controlling an arbitrary phase rotation, while reducing the Trotter error.}
    \label{fig:double_phase}
\end{figure}
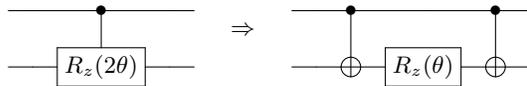

\item \textbf{Double-phase trick:} In the Hadamard test circuit underlying our algorithm, we need to perform controlled time evolution. Since controlled arbitrary-angle rotations such as those appearing in the CDF approach are expensive, we opt to remove them by replacing each controlled $R_z$ rotation by an un-controlled $R_z$ rotation sandwiched with CNOTs, as shown in the circuit diagram of ~\cref{fig:double_phase}. As $\sigma_x \sigma_z \sigma_x = -\sigma_z$, this strategy effectively implements
\begin{equation}
        \begin{pmatrix}
            U^\dagger & 0\\
            0 & U
        \end{pmatrix} \quad \text{instead of} \quad         \begin{pmatrix}
            1 & 0\\
            0 & U
        \end{pmatrix}.
\end{equation}
Since controlling a rotation can be implemented via two uncontrolled rotations (see Fig.~4 a and b in~\cite{haner2018software}), the addition of a control would increase the number of Toffoli gates by a factor of 2. Thus, avoiding the control amounts to a $\times 2$ reduction in the total Toffoli-gate cost.
In addition to removing the expensive control, we also effectively 
double the number of time steps we simulate for, because $\ket{0}\bra{0}\otimes U^\dagger + \ket{1}\bra{1}\otimes U$ implements the same relative phase as $\ket{0}\bra{0}\otimes 1 + \ket{1}\bra{1}\otimes U^2$~\cite{babbush2018encoding}, while the cost stays the same. This is an additional effective Toffoli-gate cost reduction by a factor of $\times 2$.
Alternatively, we can use the ``double-phase" strategy to mitigate Trotter error. For a $k$-th order product formula, the leading-order error in eigenvalues and eigenstates scales as $O(\Delta^k)$, where $\Delta$ is the Trotter step size. By maintaining the number of steps while halving the Trotter step size, we can achieve a $2^k$-fold reduction in the Trotter error. Altogether, the double phase trick ends up reducing the cost by a combined factor of $\times 4$.

\item \textbf{Combining consecutive rotations.} The consecutive unitaries appearing in \cref{fig:CDF} from neighbouring fragments can be straightforwardly combined into a single rotation, $\bm{V}^{(\ell)}:=\bm{U}^{(\ell-1)} (\bm{U}^{(\ell)})^T$, since they correspond to one compound single-particle basis rotation $V^{(\ell)} := U^{(\ell-1)} (U^{(\ell)})^T$. This halves the number of basis rotations required per Trotter step, which reduces the per-step gate costs by approximately a third. 

\item \textbf{Rotation precision.} The $\sigma_z$-rotations needed in the CDF-based time evolution do not need to be implemented with very high precision, but only be precise enough to reconstruct the spectrum to desired accuracy.  Limiting the accuracy of rotations is equivalent to limiting the precision of the optimized terms in the CDF decomposition, further approximating the Hamiltonian. Setting rotation precision based on the final observable of interest results in lower cost than setting it based on the standard approach of counting the total number of rotation gates. 
\end{enumerate}

\subsection{Choosing the simulation time step\label{ssec:hamiltonian_simulation}}

The cost of performing time evolution using a $2k$-order product formula for time $t$ while respecting precision $\epsilon$, namely $|e^{-iHt} - U_{\text{Trot}}(t)| < \epsilon$,  is well-known and scales as $O(t^{1+\frac{1}{2k}}\epsilon^{-\frac{1}{2k}})$. Based on this scaling, one might expect a similar dependence for the time evolution required in each Hadamard test, see e.g.~\cite{lin2022heisenberg,dong2022ground,trenev2025refining}. However, in the particular context of problems aiming to compute spectral features of the Hamiltonian at fixed accuracy, e.g. in spectroscopy, a better time scaling is possible. This is enabled by the fact that what is important in the XAS application is the error in the spectrum $G_\rho(\omega)$ -- potentially a much less stringent requirement than exact agreement of the propagators.

We illustrate this on a simple example of a second-order product formula. For a Hamiltonian $\hat{H} = \sum_j H_j$ split into non-commuting fragments $H_j$ and with a time step $\Delta$, a second-order product formula approximates time evolution under the full Hamiltonian as the product of the exponentials of the individual fragments 
\begin{equation}
   e^{-i\sum_j H_j t} \approx U_2(H, \Delta) = \prod_{j=1}^N e^{-i\frac{\Delta}{2} H_j} \prod_{j=N}^1 e^{-i\frac{\Delta}{2} H_j}.
\end{equation}
Using the Baker-Campbell-Hausdorff formula~\cite{blanes2004convergence}, we can re-combine this into a single exponential
\begin{multline}
    U_2(H,\Delta)= \exp\Bigg(-i \Delta\sum_j H_j \\
    -i\Delta^3\sum_{j}  \Bigg[\frac{[H_j,[ \sum_{h<j}H_h, H_j]]}{24} \\
    + \frac{[\sum_{h<j} H_h,[ \sum_{h<j}H_h, H_j]]}{12} \Bigg]
   +\ldots \Bigg).
\end{multline} 
From these equations, we see that rather than approximating the time evolution, we can view the product formula as implementing \textit{exact time evolution under an approximate Hamiltonian}.
In particular, instead of implementing time evolution under $H$, the product formula implements time evolution under the perturbed Hamiltonian
\begin{multline}\label{eq:effective_Hamiltonian_U2}
        \hat{H}'  = \hat{H}  + \Delta^2\sum_{j} \Bigg[\frac{[H_j,[ \sum_{h<j}H_h, H_j]]}{24} + \\\frac{[\sum_{h<j} H_h,[ \sum_{h<j}H_h, H_j]]}{12} \Bigg] + \dots,
\end{multline}
where we view the leading term in the expansion 
\begin{multline}\label{eq:Y3}
 Y_3:=-\sum_{j} \Bigg[\frac{[H_j,[ \sum_{h<j}H_h, H_j]]}{24}\\ + \frac{[\sum_{h<j} H_h,[ \sum_{h<j}H_h, H_j]]}{12} \Bigg]
\end{multline}
as a perturbation. We then are free to choose $\Delta$ sufficiently small to make the shift in eigenvalues and eigenvectors due to the perturbation $Y_{3}$ smaller than the respective target errors for those quantities.
As $\Delta$ is typically small, we can use standard perturbation theory to estimate the effects of $Y_3$
on the spectrum.
Denoting by $(E_l, \ket{E_l})$ the exact eigenvalues and eigenvectors of $H$, and by $(E'_l, \ket{E'_l})$ those of $\hat{H}' $, we obtain
\begin{align}\label{eq:perturbation_eigenstate}
    \ket{E'_l} &= \ket{E_l} - \Delta^{2} \sum_{k\neq l} \frac{\braket{E_k|Y_{3}|E_l}}{E_k-E_l}
    \ket{E_k}+O(\Delta^{4}),\\
    \label{eq:perturbation_eigenvalues}
    E'_l &= E_l - \Delta^{2}\braket{E_l|Y_{3}|E_l} + O(\Delta^{4}).
\end{align}

From these equations, it suffices to set a value of $\Delta$ such that the eigenvalue and eigenvector shift is within target accuracy.  We do not necessarily need to reduce the error in the full time signal, i.e., control the general error in the evolution operator $|e^{-iHt} - U_{\text{Trot}}(t)| < \epsilon$. Once we have found a suitable $\Delta$ to compute the spectrum sufficiently accurately, we can implement as many steps as necessary from the product formula, independently of the time signal error. That is, at fixed target precision \textit{there is no need to take shorter steps for longer simulation times}. Consequently, in each Hadamard test the cost will only scale linearly with time. 
More explicitly, once we fix $\Delta$ through \cref{eq:perturbation_eigenvalues}, we may choose $\tau j / \Delta$ as the number of Trotter steps needed to simulate to time $\tau j$, rather than the steeper scaling number $(\tau j/\Delta)^{1+1/2k}$ implied by the usual Trotter analysis for a product formula of order $2k$, resulting in lower overall cost.
Similar arguments apply to any algorithm aiming to study the spectrum of a Hamiltonian whose evolution is implemented via a product formula. In this work, we determine $\Delta$ by focusing on the peak position shift (\cref{eq:perturbation_eigenvalues}). While controlling the changes to the eigenstates (and thus peak intensities) is also of general interest, in this work we focus on the peak shifts and leave the intensity changes control to future work. This is motivated by the fact that peak shifts appear to be more consequential to the spectral fingerprinting task than intensity changes, while also being simpler to evaluate.

\subsection{Sampling distribution}
\label{ssec:distributed-sample}

Our final key cost reduction is implementing an efficient sampling strategy. A cost-efficient approach for computing the Green's function in \cref{eq:green-via-discretetime-fourier} is to combine the time evolution times $j$ and the shots needed to estimate the expectation value simultaneously in a Monte Carlo sampling scheme, as first described in Ref.~\cite{lin2022heisenberg} for ground-state energy estimation and proposed for spectroscopy in Ref.~\cite{fomichev2024simulating}. The key motivation to do this is to take advantage of the exponentially decaying Lorentzian kernel $e^{-\eta\tau |j|}$. The kernel makes longer times contribute progressively less to the overall spectrum: devoting the same number of samples to all evolution times would be wasted on trying to accurately estimate the long-time expectation values whose overall contribution is heavily suppressed. Said differently, a kernel-aware sample allocation can achieve reasonable accuracy with much fewer samples.

We define a random variable $J$ sampled from the Lorentzian kernel $e^{-\eta\tau|J|}$, taking (integer) values in a fixed maximum evolution time range $[-j_{\text{max}}, j_{\text{max}}]$ . We replace the sum over evolution times $j$ in \cref{eq:green-via-discretetime-fourier} and the statistical sampling by a single, explicit expectation value with respect to sampling $J$. With a total samples (shots) budget $S$, we estimate the Green's function as
\begin{multline}\label{eq:green-sampling}
    - \eta \operatorname{Im} G_\rho(\omega) = \frac{\eta\tau}{2\pi} \frac{A}{S} \sum_J\Big(\operatorname{Re}\tilde{G}(\tau J) \\
    + i\operatorname{Im}\tilde{G}(\tau J) \Big) e^{iJ\tau\omega},
\end{multline}
where $A = \sum_{j=-j_{\text{max}}}^{j_{\text{max}}} e^{-\eta\tau |j|}$ is the Lorentzian kernel normalization constant and the sum is over a total of $S$ instances of drawing the random variable $J$ from the range $[-j_{\text{max}}, j_{\text{max}}]$. 
Since the terms in \cref{eq:green-via-discretetime-fourier} with $j < 0$ are complex conjugate counterparts to those with $j > 0$, we can simplify the above expression for the spectrum, arriving at the following form
\begin{multline}\label{eq:greens-final}
    - \eta \operatorname{Im} G_\rho(\omega) = \frac{\eta\tau}{2\pi} \frac{A}{S} \Big( 1 + 2\sum_{j=1}^{j_{\max}} S_j\Big[ \\
    \cos(\tau j \omega) \mathbb{E}\left(\operatorname{Re}\tilde{G}(\tau j)\right) \\
    -\sin(\tau j \omega) \mathbb{E}\left(\operatorname{Im}\tilde{G}(\tau j)\right) \Big] \Big),
\end{multline}
where the expectation values of the real and imaginary parts of $\tilde{G}(\tau j)$ are performed with exactly $S_j$ shots from the overall budget. For the sake of simplifying the analysis, we have taken the number of samples at time $j$ to correspond directly to the fraction of the total shots budget dictated by the value of $e^{-\eta\tau j}$ (which is true on average when sampling $J$ according to $e^{-\eta \tau |j|}$). Specifically, for a total shots budget $S$, and normalization constant $A$, 
each time $j$ is assigned $S_j = S e^{-\alpha \eta\tau |j|} / A$ shots. Taking $\alpha = 1$ assigns a number of shots directly proportional to the Lorentzian kernel, minimizing the total number of shots required to achieve a fixed target precision. In practice we know that an $\alpha$ slightly larger than 1 will be preferable because shorter time evolution circuits are cheaper. Overall, this sampling strategy saves significant resources compared to distributing samples uniformly among different times $j$. 

\begin{figure}[t]
    \centering
    \[
    \Qcircuit @C=1em @R=0.7em {
        \lstick{\ket{0}} & \qw & \gate{H}  & \gate{S^\dagger} & \ctrl{1} \gategroup{1}{4}{1}{4}{.7em}{--} & \gate{H}  & \meter & \cw& \cw   \\
        \lstick{\ket{0}} & {/}\qw & \gate{U_\rho} & \qw & \gate{U} & \qw & \gate{U_\rho^\dagger}  & \meter & \cw\\
    }
    \]
    \caption{Double measurement circuit, replacing~\cref{fig:algo-circs-timedomain}, where $U = \exp(-i H\tau j)$ and $U_\rho$ represents the (unitary) state preparation of state $\frac{\hat{m}_{\rho}\ket{I}}{\|\hat{m}_{\rho}\ket{I}\|}$. After the Hadamard test, the state in the system register is $\propto (1\pm U)U_\rho\ket{0}$ or $\propto (1\pm i U) U_\rho\ket{0}$, depending on whether we were measuring the real or imaginary component of $U$, based on the absence or presence of the $S^\dagger$ gate.
    If $U_\rho$ is unitary, we can reverse the state preparation in the system register and measure the probability of projection into $\ket{0}$, which will be proportional to $|\braket{\rho|(1+i^p U)|\rho}|^2$ for $p\in\{0,\ldots 3\}$. We thus get two measurements with just one implementation of (the controlled version of) $U$.
    }
    \label{fig:main_text_double_measurement}
\end{figure}
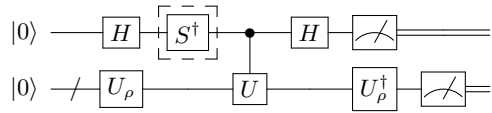
\subsection{Double measurement trick}\label{ssec:double_measurement}

Finally, we present one last algorithmic technique that we refer to as the ``double measurement trick''. In the Hadamard test, it is standard to discard the state left in the system register after time evolution, as only the ancilla measurement result is used. However, there is extra information in that state. One way to make use of it is to evaluate the probability that this output state projects back into the initial state $\ket{\rho} = \hat m_\rho\ket{I} / \|\hat m_\rho\ket{I}\|$. This makes sense to do whenever preparing the initial state $\ket{\rho}$ is significantly cheaper than the time evolution dictated by $U$. Assuming the state preparation is unitary -- as is the case for the sum-of-Slaters method~\cite{fomichev2024initial} -- performing this projection requires no extra qubits and only adds the cost of the initial state un-preparation, see~\cref{fig:main_text_double_measurement}. (A non-unitary version can be executed with the SWAP test, see ~\cref{fig:double_measurement} in \cref{appsec:double_measurement_trick}.) Since the unnormalized output of the Hadamard test is proportional to $(1+i^p U)\ket{\rho}$ for $p\in\{0,\ldots,3\}$, the second measurement will allow us to measure $|\braket{\rho|(1+ i^p U)|\rho}|^2$ from which we can estimate the value of $\braket{\rho|U|\rho}$ of interest. 

The amount of information we can derive with this trick will be dependent on $\braket{\rho|U|\rho}$. Empirically we find that on average the double measurement trick provides a $\times 1.49$ multiplicative factor saving whenever the state preparation cost is subdominant. A more detailed analysis of this technique is provided  in~\cref{appsec:double_measurement_trick}.

\section{Resource estimation}
\label{sec:resources}

Having described the algorithm, its implementation, and the key optimizations, here we perform the constant factor resource analysis. We present both the formulas for computing the total cost of the algorithm in terms of logical qubits and Toffoli-gate count, as well as the error analysis required to fix the free parameters in those formulas. We conclude this section by describing how the algorithm could be compiled into the active volume architecture. 

\subsection{Cost formulas}

At the highest level, the Toffoli-gate count of the algorithm can be written as
\begin{equation}
    N_T = N_{\text{Trot}} \times C_{\text{Trot}},
    \label{eq:global-cost}
\end{equation}
where $N_{\text{Trot}}$ is the total number of individual Trotter steps required to execute the full algorithm, and $C_{\text{Trot}}$ is the cost of each Trotter step. The gate cost of initial state preparation using the sum-of-Slaters method \cite{fomichev2024initial} is significantly smaller than that of time evolution, $2.75\times 10^5$ gates for $D = 10^4$ Slaters.

\textbf{Number of Trotter steps:} 
We first estimate the total number of Trotter steps required $N_{\text{Trot}}$. Given a shot budget $S$, using the sampling distribution strategy as discussed in \cref{ssec:distributed-sample} we assign $S_j$ shots to computing $G(\tau j)$, where
\begin{equation}\label{eq:shot_distribution}
    S_j = S\frac{\exp(-\alpha \tau \eta |j|)}{A},\qquad  A = \sum_{j=1}^{j_{\max}} e^{-\alpha |j|\tau \eta},
\end{equation}
and $\alpha$ is an optimization parameter.
For each $G(\tau j)$ we have to simulate $\tau j / \Delta$ Trotter steps with time step $\Delta$, as argued in \cref{ssec:hamiltonian_simulation}.
 Using Hadamard tests to measure the real and imaginary components along three Cartesian axes, and summing over all the shots $S_j$ to be performed at each time $\tau j$, the total number of Trotter steps to be taken in the entire algorithm is then
\begin{align}\label{eq:N2T}
        N_{\text{Trot}} 
        &= 3 \cdot 2\cdot S \cdot \frac{\tau}{\Delta} \sum_{j=1}^{2j_{\text{max}}} \frac{ j  \exp{(-\alpha\tau\eta |j|)}}{A},
\end{align}
where the total shot budget $S$, maximal evolution time $j_{\max}$, the scaling $\tau$, and the Trotter time step $\Delta$ are yet to be chosen.
Meanwhile the longest circuit is independent of the number of shots and only requires
\begin{align}\label{eq:N2Tmax}
        N_{\text{Trot},\max} 
        &= j_{\max}\frac{\tau}{\Delta} 
\end{align}
Trotter steps. The factor $\tau / \Delta$ indicates the number of Trotter steps per sampling time step. These equations display the linear cost scaling in time consequence of the perturbative analysis of the spectrum discussed in~\cref{ssec:hamiltonian_simulation}.

\textbf{Cost of a Trotter step:} The cost of a Trotter step $C_{\text{Trot}}$ will be determined by the number of calls $N_{\text{Trot calls}}$ to the first-order Trotter formula (for higher-order formulas), the number of fragments $L$ in the CDF of the Hamiltonian, and the cost of implementing time evolution of each fragment $C_{\text{fragment}}$,
\begin{equation}
    C_{\text{Trot}} = N_{\text{Trot calls}} \times L \times C_{\text{fragment}}.
    \label{eq:perstep-cost}
\end{equation}
The number of calls to first-order Trotter is straightforwardly determined by the choice of the product formula, while $L$ is another free parameter to be determined by the need to control the CDF approximation error. 

To compute the cost of time evolution for a single fragment $C_{\text{fragment}}$, we note that for each fragment indexed by $\ell$, we need to implement (i) the unitary $\bm{U}^{(\ell)}$ associated to the single-particle basis rotation $U^{(\ell)}$ and (ii) the corresponding term $Z_{kl}^{(\ell)}$. Note that only one basis rotation per Trotter step is required as discussed at the end of \cref{ssec:cdf}.
The basis rotation requires $2\frac{N(N-1)}{2}$ Givens rotations, each of which can be implemented with one $X\otimes X$ plus one $Y\otimes Y$ rotation, and $2N$ single-qubit $Z$ rotations \cite{arrazola2022universal}. Each $Z_{kl}^{(\ell)}$ requires $\frac{2N(2N+1)}{2}$ $Z\otimes Z$ rotations, except for the one-body $Z^{(0)}$, which only needs $2N$ $Z$ rotations. Further, each $Z\otimes Z$ rotation can be implemented with a single $Z$ rotation and two CNOT gates, as  depicted in~\cref{fig:rzz}.
Each single qubit rotation can be implemented via phase kickback using the qubit as a control qubit for the addition on the binary expression of the angle on a phase gradient state~\cite{gidney2018halving},
\begin{equation}
    \ket{\text{Grad}_n} = \frac{1}{\sqrt{2^n}}\sum_{k=0}^{2^n-1}  e^{2i\pi k/2^n}\ket{k}.
\end{equation}
Using the adder subroutine introduced in~\cite{gidney2018halving}, this results in \begin{equation}
    C_{\text{rot}}(\epsilon_{\text{rot}}) = \lceil\log_2(\epsilon_{\text{rot}}^{-1})\rceil
\end{equation}
Toffoli gates. Combining all of this, the total Toffoli-gate cost of implementing time evolution by a CDF fragment is
\begin{equation}
    C_{\text{fragment}} = C_{\text{unitary}} + C_{\text{Z-matr}}, 
\end{equation}
with
\begin{align}
    C_{\text{unitary}} &= 2 \frac{N(N-1)}{2} \left(2 C_{\text{rot}}\right)   + 2N C_{\text{rot}}, \label{eq:unitary-cost}\\
    C_{\text{Z-matr}} &=  \frac{2N(2N-1)}{2} C_{\text{rot}} \quad \text{or} \quad 2N C_{\text{rot}}, \label{eq:zrot-cost}
\end{align}
the latter depending on whether the fragment is two-body or one-body ($\ell = 0$), respectively.

Computing the logical qubit cost of the algorithm is straightforward: we require $2N$ qubits for the system register, $1$ control for the Hadamard test, and $5\log_2(D)-3$ additional qubits for the preparing the initial state using the sum-of-Slaters procedure \cite{fomichev2024initial}. The qubit used as a control for the Hadamard test can be one of the state preparation auxiliary qubits. Thus, we have,
\begin{equation}
    N_{\text{qubits}} = 2N + 5\log_2(D) - 3.
\end{equation}

Using the above equations we can directly compute the total Toffoli-gate count of the algorithm once we fix the free parameters: total shots budget $S$, maximal evolution time $j_{\text{max}}$, the scaling $\tau$, the Trotter time step $\Delta$, the number of fragments $L$ in the CDF, and the rotation precision $\epsilon_{\text{rot}}$ ($\eta$ is fixed by the experimental broadening of $1$ eV). In the following section we use error analysis to constrain the choice of these parameters.

\subsection{Fixing free parameters through error analysis}\label{ssec:fixing_parameters}

The choice of the free parameters $S, j_{\text{max}}, \tau, \Delta, L$, and $\epsilon_{\text{rot}}$ is determined by six sources of error. Three of them --- the number of fragments $L$, rotation precision $\epsilon_{\text{rot}}$ and Trotter error controlled by $\Delta$ --- represent the effect of various approximations to the Hamiltonian. Two of them, $j_{\text{max}}$ and $\tau$, are due to the truncation of the formally infinite sum in \cref{eq:green-via-discretetime-fourier}, and discretization, i.e. the use of the discrete-time Fourier transform to represent the continuous spectral signal $G_\rho(\omega)$, respectively. The last one, the number of shots $S$, is related to the statistical sampling noise associated with collecting the time signal. We will now discuss them one by one.

\begin{enumerate}
    \item \textbf{Number of fragments in the CDF:} The number of fragments $L$ needed to achieve a fixed accuracy scales approximately linearly with the number of orbitals $L = \tilde{O}(N)$~\cite{motta2021low,von2021quantum,lee2021even}. Our empirical results, shown in ~\cref{fig:CDF-error-LiMnO} in \cref{app:CDF}, also validate this scaling. Based on these results, we estimate that we need to choose rank $L = N$ in~\cref{eq:(pq|rs)} to achieve the target precision of $1$ eV --- approximately $0.039$ Ha. Overall, thanks to the $L = N$ relationship, compressed double factorization achieves a $\tilde{O}(N^3)$ cost growth, in contrast to the $O(N^4)$ growth of the direct Pauli-based approach.

    \item \textbf{Trotter time step:} The error source determining the choice of the Trotter time step $\Delta$ is the Trotter error. As derived in ~\cref{eq:perturbation_eigenvalues,eq:perturbation_eigenstate} the error in the eigenvalues is
\begin{align}
    \epsilon_{\text{trot}} = \max_l \Delta^{2k}\braket{E_l|Y_{2k+1}|E_l},
\end{align}
to first order in perturbation theory, for a $2k$ order product formula whose leading order term in the effective Hamiltonian is $Y_{2k+1}$, see~\cref{eq:effective_Hamiltonian_U2,eq:Y3}.
Our goal is to ensure this error remains below $1$ eV, the experimental broadening of XAS spectra, so we need to select
\begin{align}
    \Delta \leq \left(\frac{|\langle Y_{2k+1} \rangle |}{1\text{ eV}}\right)^{1/{2k}},
\end{align}
    where the expectation value is with respect to a state contributing to the XAS spectrum that generates the largest error. One may use low-bond-dimension matrix-product  state approximations to the eigenstates of the Hamiltonian obtained with the density-matrix renormalization group method, and compute expectation values by representing the error operator $Y_{2k+1}$ with a matrix-product operator and performing a contraction. In this paper, we estimate the Trotter error comparing the linearized time evolution of one short step of time length $\Delta$ vs $n$ steps of time length $\Delta/n$, as discussed in~\cref{ssec:cost}. It is also worth highlighting that here we would be bounding only the error in the intensity of the peaks, but there is also a shift in the position of the peaks due to the perturbation of eigenstates, which we do not aim to bound in this paper.

    \item \textbf{Evolution time and time step:} Two more error sources are related to the truncation and discretization of the Fourier transform integral. They are associated to variables $j_{\max}$ and $\tau$ respectively. The error $\epsilon_{\text{trunc}}$ arises from truncating the formally infinite boundaries of the Fourier transform integral to a maximal evolution time $j_{\text{max}}$: it can be bounded as
    \begin{multline}
        \epsilon_{\text{trunc}} \leq \frac{2\eta\tau}{2\pi} \int_{j_{\max}}^\infty e^{-|j|\tau \eta} \\
        \times |e^{-ij\tau w}| \, \|\braket{\rho|e^{-iH\tau j }|\rho}\| dj.
    \end{multline}
    See the detailed derivation in~\cref{appssec:truncation_error}. Meanwhile the discretization error $\epsilon_{\text{disc}}$  arises from using the discrete-time Fourier transform on a continuous spectral signal $G$, as we do in~\cref{eq:green-via-discretetime-fourier}. This error may be estimated by considering a Riemann sum approximation to the exact Fourier integral over $[0, 2j_{\max}\tau]$: assuming the midpoint rule and time step $\tau$, we find    \begin{align}\label{eq:Riemann_sum_error}
        \epsilon_{\text{disc}} = \frac{\eta }{2\pi}\left\vert\int_0^{j_{\max}\tau} f(t)\, dt - \tau\sum_{j=0}^{j_{\max}} f\left(\tau j\right)\right\vert\\
        \le \frac{\eta(j_{\max}\tau)^3 \max_{\xi\in(0,j_{\max}\tau)}|f''(\xi)|}{24\pi j_{\max}^2}, \nonumber
    \end{align}
    where $f(t) =G(t) e^{it(w + i\eta)}$.
    These two equations provide relationships that $\tau$ and $j_{\max}$ should satisfy to control the truncation and discretization error. The equations can be partially inverted to constrain the values of the free parameters $j_{\text{max}}$ and $\tau$, namely 
\begin{align}
    j_{\max} &= \frac{\pi}{2\eta \tau} \log\left(\frac{1}{\epsilon_{\text{trunc}}}\right) ,\\
    j_{\max}\tau^3 &=\frac{24\pi \epsilon_{\text{disc}}}{\max_\omega|H-\omega-i\eta|^2}\frac{1}{\eta}, \label{eq:epsilon_disc}
\end{align}
for $\omega$ a frequency in the window over which the spectrum of the initial state will have support (see~\cref{appssec:truncation_error} for more details). 

If an eigenstate $\ket{E_F}$ is optically inactive, meaning that $\bra{E_F} \hat m_\rho \ket{E_I} = 0$, its contribution to the full Green's function vanishes both in the frequency and in the time domain. That means there is no need to rescale the Hamiltonian by its full spectral norm $\|H\|$, as is normally done to ensure all eigenvalues fall within $[-\pi,\pi]$ to prevent aliasing. Instead, it is sufficient to rescale the Hamiltonian such that only the \textit{optically active} eigenstates are within $[-\pi,\pi]$: any optically inactive eigenstates outside this range will not contribute anything to $\tilde G(\tau j)$ and thus will not introduce any aliasing. This is an important advantage of a Trotter-based approach to time evolution relative to a method that relies on a block-encoding, such as qubitization: in those cases one always has to rescale by the full spectral norm of the Hamiltonian, often leading to higher costs. From the above, we define
$\|H\|_\omega:=\max_\omega|H-\omega-i\eta|^2$ as a normalizing factor of the time scales for the optically active eigenstates. From these relations, we have $j_{\max} = \tilde{O}((\eta\tau)^{-1}) = \tilde{O}(\|H\|_\omega)$. 

    \item \textbf{Shots budget:} The last source of error is the measurement noise in each Hadamard test, determining the required shot budget $S$. Let us distribute the shot budget according to~\cref{eq:shot_distribution}.
    Then, the number of shots needed to achieve a measurement error $\epsilon_{\text{meas}}$ is
    \begin{equation}
        S = \left(\frac{\eta\tau}{2\pi\epsilon_{\text{meas}}}\right)^2 \sum_j e^{(\alpha-2)|j|\tau\eta}\sum_k e^{-\alpha|k|\tau\eta},
    \end{equation}
see~\cref{appssec:measurement_noise}. Meanwhile the uniform distribution ($\alpha = 0$) would result in a number of shots
\begin{equation}
    S_{\text{uniform}}  = \frac{j_{\max}}{2} \sum_j e^{-2 j\eta \tau}\left(\frac{\eta\tau}{2\pi\epsilon}\right)^2.
\end{equation}
For a concrete choice of the variables $j_{\text{max}}, \tau, \eta$, these expressions may be used to estimate the reduction in sampling costs obtained with the proposed sampling strategy. 

\end{enumerate}

\subsection{Active volume\label{sssec:active_volume}}

To complete our resource estimation analysis, and to allow us to make rough estimates of runtime for our quantum algorithm in \cref{sec:application}, we analyze the cost of compiling our circuit onto an active volume architecture~\cite{litinski2022active}. This architecture divides the qubit space into memory and workspace regions, each with logarithmic connectivity, enabling efficient routing of surface code patches~\cite{litinski2019game}. By employing a form of gate teleportation known as backwards-in-time idling (see~\cite[Fig. 10]{litinski2022active}), the architecture significantly parallelizes the implementation of many gates, including non-commuting ones, thereby reducing circuit depth.

In Ref.~\cite{litinski2022active}, Table I provides a compilation of common quantum circuit subroutines, including single-qubit rotations, which are the most expensive operations in our CDF-based Trotter simulation. Based on this table and a list of the key operations, one can compute the active volume $V$ for the circuit. The active volume and the total number of available logical qubits $n_q$, both in memory and workspace combined, determines the circuit depth,
\begin{equation}\label{eq:active-volume}
    \text{circuit depth} \gtrsim \frac{2V}{n_q}.
\end{equation}
We shall use the expressions in Table I of Ref.~\cite{litinski2022active} and \cref{eq:active-volume} to compute the active volume and circuit depth for our chosen target system in \cref{sec:application}. 
Using the circuit depth and choosing a clock rate for the quantum computer, we can estimate the runtime required to execute a circuit. For example, if a circuit has active volume  $V=10^{10}$ and we have a quantum computer with $n_q = 200$ logical qubits  that runs at 1 MHz clock rate, we will need $2\cdot 10^{10} / (200 \cdot 10^{6}\text{ Hz} )= 100$ seconds to execute the circuit.

\section{Application: Lithium-excess Cathodes}
\label{sec:application}
We now apply our highly optimized algorithm to an industrially relevant system -- the Li$_4$Mn$_2$O cluster with the CAS(22e,18o) active space, a prototypical minimal system for studying structural degradation in a Li-excess-based battery cathode. 
First, we implement and run the algorithm as described in \cref{sec:optimizations} on a simulator, and demonstrate by comparing against a classical reference that it correctly predicts the absorption spectrum. We then use the analysis from \cref{sec:resources} to estimate constant-factor resource requirements for the Li$_4$Mn$_2$O model system, documenting improvements from each of the optimizations. We finish by providing a rough comparison of runtimes between our quantum algorithm and a state-of-the-art classical method for simulating XAS, the restricted active space (RAS) approach. 

\subsection{Li-excess cluster system}

We follow the strategy described in Ref.~\cite{fomichev2024simulating} to extract a prototypical oxygen-centered cluster Li$_4$Mn$_2$O for studying structural degradation of Li-excess cathodes. For this cluster, we employ the cc-pVDZ basis set, obtain Hartree-Fock molecular orbitals, and from those construct a minimal (valence) active space, consisting of the ten $3d$ orbitals of the Mn atoms, the four $2s$ valence orbitals of the Li and the three $2p$ O orbitals, together with the core $1s$ O orbital. This gives a target active space of $N = 18$ spatial orbitals, though in~\cref{tab:resources} we also consider the cost of adding the $3p$ or the $4d$ orbitals of Mn. The active space is built using the atomic valence active space method (AVAS \cite{sayfutyarova2017automated}) as implemented in the code PySCF \cite{sun2015libcint,sun2018pyscf,sun2020recent}. Within this active space, we used tools in PySCF to create both the Hamiltonian $\hat H$ and the dipole operator $\hat m_\rho$. To determine the initial state $\hat m_\rho \ket{I}$, we used the complete active space (CAS) method as implemented in PySCF to find the ground state $\ket{I}$, then applied the dipole operator using a custom implementation of the associated one-body creation-annihilation operator pairs. 

\subsection{Simulating the algorithm}
\label{ssec:simulation}

We simulate the circuit depicted in \cref{fig:algo-circs-timedomain} using the implementation described in \cref{sec:optimizations}. The implementation is done in the quantum computing software library PennyLane~\cite{bergholm2018pennylane}, and we use the \texttt{lightning} backend simulator \cite{asadi2024} to run the algorithm. Plugging the Hadamard test measurement outcomes into \cref{eq:greens-final} allows us to reconstruct the absorption spectrum. 

As a proof-of-principle of the capabilities of the algorithm to simulate absorption spectra, we first benchmark it on the valence-excitation spectrum of the N$_2$ molecule, where we can compare the result with the spectrum obtained through CAS calculations. We employ the STO-3G basis set, set the bond length to $1.077$ \AA, and work in the full orbital space of CAS(14e, 10o): this amounts to a 20+1 qubit simulation. To establish a classical reference, we performed full configuration interaction (FCI) within the active space to determine 250 lowest-energy eigenvalues and eigenstates, then used the eigenstates to calculate the matrix elements of the dipole operator. The full classical reference spectrum is then constructed using these energies and matrix elements, following ~\cref{eq:crosssection}. The quantum simulation is performed with maximal evolution time $j_{\text{max}} = 50$, shots budget $S = 10^4$, the Hamiltonian rescaling parameter $\|H\|_\omega = 3$ Ha, sampling time step $\tau = \pi / 2\|H\|_\omega$, Trotter step $\Delta = \tau$ and using the randomized second-order Trotter product formula. 

\begin{figure}[t]
    \centering    \includegraphics[width=0.8\linewidth]{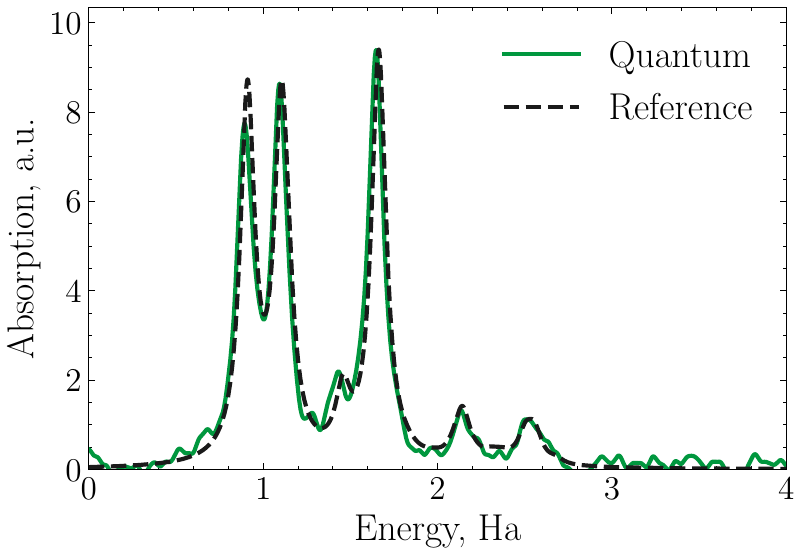}
    \caption{Benchmarking the XAS algorithm: valence-excited states of N$_2$. The simulation parameters are $j_{\text{max}} = 100$, $S = 10000$, $\|H\|_{\omega} = 3$ Ha, $\Delta = \tau = \pi/2\|H\|_\omega$, and we used the randomized second-order Trotter product formula. The classical reference is obtained with the full configuration interaction method. The quantum algorithm accurately reproduces all key features of  the absorption spectrum. 
    }
    \label{fig:benchmark_n2}
    \centering
    \end{figure}

The results, shown in \cref{fig:benchmark_n2}, demonstrate that the algorithm accurately recovers the absorption spectrum of the N$_2$ molecule. While there are minor fluctuations due to the finite number of shots and a truncated maximal evolution time, all key features of the spectrum are captured at a sufficient accuracy for spectral fingerprinting. 
While the two spectra agree closely, the approach taken to arrive at the result is drastically different. In particular, the classical method calculation required us to calculate a very large number of eigenstates to be able to reconstruct the spectrum. More than just a nuisance, this is a fundamental problem with the state-by-state approach of computing the absorption spectrum employed by many classical methods: as the size of the system grows, the number of states that need to be converged to resolve all key aspects of the spectrum becomes prohibitively large. Instead, by performing the calculation in the time-domain, for which it is uniquely suited, the quantum algorithm effectively obtains the entire spectrum at once -- even for the shortest possible simulation times. Subsequent effort of doing longer-time evolutions is then focused on refining and resolving finer spectral features. This is very different than simply having no data at all in the higher-energy region of interest until one had churned through many lower-energy eigenstates, as is the case with standard classical methods. The density of states will only increase with the addition of more orbitals, easily reaching hundreds or thousands of states in the XAS region, meaning that the state-by-state approach will quickly become infeasible.

Having benchmarked the algorithm on the example of N$_2$ where we could still access the absorption spectrum classically, we then simulate the spectrum for the industrially relevant Li$_4$Mn$_2$O cluster. We use a smaller active space of only the ten Mn $3d$ orbitals together with the core O $1s$ orbital, labeled CAS(14e,11o), resulting in a simulation of 22+1 qubits. The key difference with the N$_2$ case is that here we simulate core-excited states that actually make up the XAS spectrum, rather than the valence-excited spectrum we used for benchmarking with N$_2$. 

\begin{figure}[t]
    \includegraphics[width=0.8\linewidth]{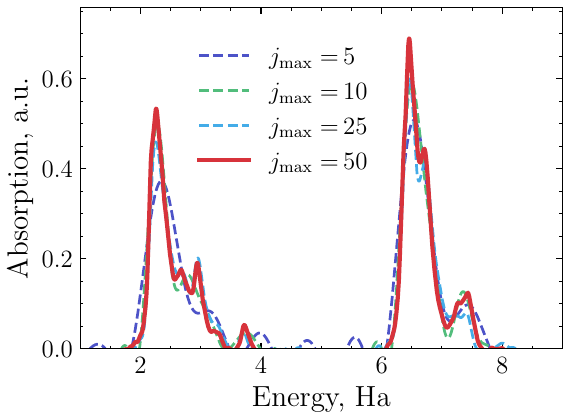}
    \caption{Benchmarking the XAS algorithm: core-excited states of the LiMnO cluster. In lieu of a comparison with a classical reference, we show convergence of the spectrum as a function of the maximal evolution time $j_{\text{max}}$. The other simulation parameters are $S= 5000$, $\|H\|_\omega = 2$ Ha, $\Delta = \tau = \pi / 2\|H\|_\omega$, and we use the second-order Trotter product formula.}
    \label{fig:benchmark_limnocluster}
\end{figure}

To be able to target only the core-excited states and save computational effort, we employed the core-valence separation (CVS) approximation \cite{cederbaum1980many,barth1981many,norman2018simulating,herbst2020quantifying}. This technique is based on the observation that electrons in molecules naturally form these two groups of core and valence, and the interactions between them, as mediated by the two-electron term in the Hamiltonian, are typically rather small. By zeroing out those small two-electron integrals, CVS allows to effectively decouple the core-excited and valence-excited many-body state manifolds. The computational cost reduction of CVS for classical methods such as RAS is limited. This is because RAS is fundamentally a state-by-state computational method, meaning that the lower-lying valence-excited states still need to be computed first, even if that manifold is decoupled from the higher-lying core-excited state manifold through the use of CVS. However, for the quantum algorithm, as long as the initial state can be placed in the core-excited state region, the time evolution will only proceed in the core-excited state manifold, reducing the effective size of the Hilbert space and thus the cost of the algorithm. To localize support of the initial state into the core-excited state manifold, we removed all dipole operator elements that do not involve a core excitation.

With the CVS modifying the initial state and the Hamiltonian, we ran the same circuit implementation as for the N$_2$ case above. The result is shown in \cref{fig:benchmark_limnocluster}. In the absence of a reference, we instead varied the parameters in the simulation until the spectrum we obtained was reasonably well-converged. The final parameter choices were $j_{\text{max}} = 50$, shots budget $S = 5000$, the Hamiltonian rescaling parameter $\|H\|_\omega = 2$ Ha, sampling time step $\tau = \pi / 2\|H\|_\omega$, Trotter step $\Delta = \tau$, and we again used the second-order Trotter product formula. The convergence as a function of maximal evolution time $j_{\text{max}}$ is illustrated with a few representative curves in \cref{fig:benchmark_limnocluster}. This simulation of the smaller version of the full CAS(22e, 18o) of Li$_4$Mn$_2$O helps verify that the XAS absorption algorithm proposed here is indeed capable of accessing the XAS response of a realistic system like Li$_4$Mn$_2$O.

\begin{table*}
\centering
\setlength{\tabcolsep}{10pt} 
\renewcommand{\arraystretch}{1.2}
\begin{tabular}{c c c c c c}
\multicolumn{2}{c}{} & \multicolumn{2}{c}{Algorithm} & \multicolumn{2}{c}{Largest Circuit} \\
\cline{3-4} \cline{5-6}
 N & Logical qubits & Toffoli gates & Active Volume & Toffoli gates & Active Volume \\
\hline
$6$ & $76$ & $2.63 \times 10^{10}$ & $9.59 \times 10^{11}$ & $1.53 \times 10^{7}$ & $6.40 \times 10^{8}$ \\
$9$ & $82$ & $6.94 \times 10^{10}$ & $3.28 \times 10^{12}$ & $4.49 \times 10^{7}$ & $2.23 \times 10^{9}$ \\
$10$ & $84$ & $9.27 \times 10^{10}$ & $4.54 \times 10^{12}$ & $6.08 \times 10^{7}$ & $3.09 \times 10^{9}$ \\
$11$ & $86$ & $1.21 \times 10^{11}$ & $6.08 \times 10^{12}$ & $8.05 \times 10^{7}$ & $4.15 \times 10^{9}$ \\
$14$ & $92$ & $2.45 \times 10^{11}$ & $1.28 \times 10^{13}$ & $1.66 \times 10^{8}$ & $8.74 \times 10^{9}$ \\
$18$ & $100$ & $5.20 \times 10^{11}$ & $2.76 \times 10^{13}$ & $3.54 \times 10^{8}$ & $1.89 \times 10^{10}$ \\
$22$ & $108$ & $9.53 \times 10^{11}$ & $5.09 \times 10^{13}$ & $6.51 \times 10^{8}$ & $3.49 \times 10^{10}$ \\
$24$ & $112$ & $1.24 \times 10^{12}$ & $6.64 \times 10^{13}$ & $8.48 \times 10^{8}$ & $4.55 \times 10^{10}$ \\
$28$ & $120$ & $1.98 \times 10^{12}$ & $1.06 \times 10^{14}$ & $1.35 \times 10^{9}$ & $7.28 \times 10^{10}$ \\
\hline
\end{tabular}
\caption{Estimates of resources required to implement the time domain XAS algorithm presented in this paper for the Li$_4$Mn$_2$O Hamiltonian in active spaces of $N$ spatial orbitals. The logical qubits indicated are the minimum number required, excluding those used as magic state factories but including the auxiliary qubits needed for initial state preparation. To obtain the number of qubits needed for state preparation using the sum-of-Slaters approach \cite{fomichev2024initial}, we assumed $D = 10^4$ number of Slater determinants in the superposition. The circuit was assumed to use a randomized second-order product formula with $\Delta = \sqrt{\eta / |\langle Y_3\rangle|}$ for $|\langle Y_3\rangle|= 1$ Ha$^3$ (see~\cref{fig:trotter_error}) and $\eta = 0.05$ Ha. We also used $\alpha =1.3384$, $\|H\|_\omega = 2$ Ha, $S = 2500$, $\tau = \pi / 2\|H\|_\omega$ and $j_{\max} = 200$, determined by constraints of the error analysis of \cref{sec:resources}. Finally, we fixed $L = N$, as suggested by~\cref{fig:CDF-error-LiMnO}, and a precision of $\epsilon_{\text{rot}} = 10^{-3}$ for each single-qubit rotation.}
\label{tab:resources}
\end{table*}

\subsection{Overall cost and impact of improvements~\label{ssec:cost}}

Having benchmarked the algorithm through simulations, we now use the methodology of \cref{sec:resources} to compute exact constant factor resource estimates for the Li$_4$Mn$_2$O model cluster with CAS(22e,18o). Our cost estimates are shown in \cref{tab:resources} for a range of active spaces for the Li$_4$Mn$_2$O cluster. In generating the numbers in the table, we set $\Delta = \sqrt{\eta / |\langle Y_3\rangle|}$ for $|\langle Y_3\rangle| = 1$ Ha$^3$ (see~\cref{fig:trotter_error} for determining the error bound) and $\eta = 0.05$ Ha, $\alpha = 1.3384$, $\|H\|_\omega = 1$ Ha, $S = 2500$, $\tau = \pi / 2\|H\|_\omega$, $j_{\max} = 200$ and $L = N$, see~\cref{fig:CDF-error-LiMnO}. To factor in the costs of initial state preparation for logical qubit count, we assume $D = 10^4$ Slater determinants are sufficient to accurately capture the initial state, and use the cost formulas in \cref{sec:resources}.
For the gate rotations we targeted $10^{-3}$ Ha precision. To use the active volume architecture we need a number of logical qubits twice as large as those reported in~\cref{tab:resources}.

Moreover, we evaluate the individual cost-reducing impact of all algorithmic optimizations we have described in \cref{sec:optimizations}, obtaining the results summarized in  \cref{fig:optimizations_and_cost}. 

\begin{enumerate}
    \item \textbf{CDF:} Using compressed double factorization with $L = N$ unlocks significant savings. A sparsified Jordan-Wigner mapping of the CAS(22e, 18o) Hamiltonian results in approximately $5.7\times 10^4$ Pauli string rotations with angles larger than $10^{-2}$ for a single first-order Trotter step with error close to $1$ eV. By contrast, the CDF Hamiltonian with rank $L = N$ only requires $2835$ Pauli string rotations, before accounting for the sparsity. This translates directly into a gate cost reduction factor of roughly $\times 20$. Note that this estimate does not take into account the Trotter error from fragmentation.

    \item \textbf{Double phase:} As argued in \cref{sec:optimizations}, replacing controlled rotations by un-controlled rotations saves a factor of 2 at the cost of two Clifford gates (CNOTs), since a controlled rotation can be implemented via 2 uncontrolled rotations; moreover, the double phase approach doubles the effective number of steps being simulated at no extra cost. Thus the overall gain is a factor of $\times4$.

    \item \textbf{Combining consecutive rotations:} Using the CDF with $L = N$ for the CAS(22e, 18o) Li$_4$Mn$_2$O cluster, we evaluate the relative costs of implementing the unitaries $\bm{U}^{(\ell)}$ versus the $Z_{kl}^{(\ell)}$ Pauli $Z$ rotations. This allows to estimate the cost savings from having one less separate unitary to apply per Trotter step, resulting in a multiplicative saving factor of $\times 1.46$.

    \item \textbf{Rotation precision}: As demonstrated in~\cref{fig:rotation-error-LiMnO}, by studying a range of active spaces for the Li$_4$Mn$_2$O cluster, we empirically find that $\epsilon_{\text{rot}} = 10^{-3}$ results in an eigenvalue error on the order of $10^{-2}$ Ha. Considering our target accuracy and other error sources, we thus choose $\epsilon_{\text{rot}} = 10^{-3}$. This represents a multiplicative factor savings of $\times 1.57$ with respect to what we would have got with the more conservative approach of assuming we need a precision $10^{-6}$ for each of the $3.5\times 10^7$ rotations in the longest circuit for the CAS(22e,18o) active space.

    \item \textbf{Perturbative analysis of the spectrum:~\label{item:perturbative}}~\cref{fig:trotter_error} shows the second and fourth order Trotter error for several chemical systems. We estimated this error by evaluating the difference between time evolutions for the same $t = 1$, but different number of Trotter steps. Using the second order product formula, we take $n = 10$ steps with $\Delta = 1 \text{ a.u.}^{-1}/ n = (10\text{ a.u.})^{-1}$; and compare against taking 4 times as many steps, $n=40$.  Explicitly:
    \begin{multline}
    |\braket{E_l|Y_{3}|E_l}| \approx \frac{1}{\Delta^2}\Big|\Big\langle E_l\Big|\Big(U_{2}^{n}(\Delta) \\
    - U_{2}^{4n}\big(\tfrac{\Delta}{4}\big)\Big)\Big|E_l\Big\rangle\Big|,
    \end{multline}
Since the Trotter error scales as $O(\Delta^2)$ in the above expression, the error in the second term will be $16$ times smaller than in the first one. 
This means that the difference we are computing captures a lion's share of the overall error inherent in the simulation with $n = 10$ steps. Similarly, for the fourth order formula we select $n = 3$ $(\Delta = 1/3)$ and compare against taking twice as many steps. (We verified the result would not change significantly if we used more steps in the second term.)

Extrapolating from the growth of this error for small systems, we conclude that it is reasonable to expect $|\braket{E_l|Y_{3}|E_l}|\leq 1$ Ha$^3$ for the CAS(22e, 18o) Li$_4$Mn$_2$O cluster. Future work should validate this thesis more robustly, and more generally understand how to accurately estimate the Trotter error for systems we cannot simulate classically.
    
Under this assumption, we select $\Delta$ such that $\Delta^2|\braket{Y_3}|\leq 0.05$ Ha, and $\tau/\Delta$ an integer. For $\tau = \pi / (4\text{ a.u.})$, this is satisfied for $\tau = 4\Delta$.

In~\cref{appssec:Trotter_error} we provide more detail about how the Trotter error translates into the spectrum.    Once the Trotter time step is fixed in this way, it does not need to be scaled down as a function of the simulation time, see~\cref{sec:resources}. Thus, the cost of a Hadamard test scales linearly in the simulation time, saving a factor of $\times 10.8$ over the whole algorithm.

    \item \textbf{Sampling distribution:} For the choice of parameters $j_{\max} = 200$ of the maximal evolution time,  $\tau = \pi / (4\text{ a.u.})$, $\eta =0.05$ Ha as above, and $N = 18$, using the expressions derived in~\cref{ssec:fixing_parameters} we find a multiplicative savings factor of $\times 17.9$ relative to uniform sampling.

    \item \textbf{Double measurement} As we explained in~\cref{ssec:double_measurement}, we can extract further information from the residual state of a Hadamard test using this double measurement approach. On average, this provides a $\times 1.49$ savings factor.
    
\end{enumerate}

\subsection{Classical vs. quantum}

In this concluding section, we use rough estimates to get a general sense of the expected performance of our quantum algorithm compared to the state-of-the-art classical method for XAS simulation: the restricted active space (RAS) approach. The estimates shown here come with a high degree of uncertainty and are the first attempt at quantifying the relative performance of classical and quantum algorithms for the XAS problem. Especially when it comes to the classical algorithm, obtaining accurate estimates would require implementing and running the RAS algorithm on an HPC system for the concrete problems of interest, which we leave to important future work.

\textbf{Quantum runtime}: To roughly estimate quantum algorithm runtime for active spaces of varying size, we use the active volume numbers in \cref{tab:resources} and convert them to total number of logical cycles, using a total qubit budget of $n_q/2$ workspace and $n_q/2$ memory qubits as in \cref{eq:active-volume}. We then assume a quantum computer with a logical clock rate of 1 MHz, and divide the number of logical cycles by the clock rate. Note that while the cost of preparing an initial state is a contributor to the overall runtime, it only accounts for at most 1\% of the overall Toffoli costs of the algorithm, as we computed in \cref{ssec:cost} using the sum-of-Slaters method of Ref. \cite{fomichev2024initial}. Moreover, the classical overhead associated with finding that initial state, for example by using DMRG, is also negligible. This is because for the range of system sizes we consider (10-30 orbitals) it is well-known that DMRG can produce high-quality ground states for active spaces with modest classical computational resources of a single workstation or at most a cluster compute node with around a hundred CPUs \cite{sharma2014low,fomichev2024initial,lee2023evaluating}. At the same time, the expensive task for DMRG is the computation of the excited-state spectrum -- precisely the reason why this work is proposing to leverage a quantum computer for this task.

\begin{figure}[t]
    \includegraphics[width=0.95\linewidth]{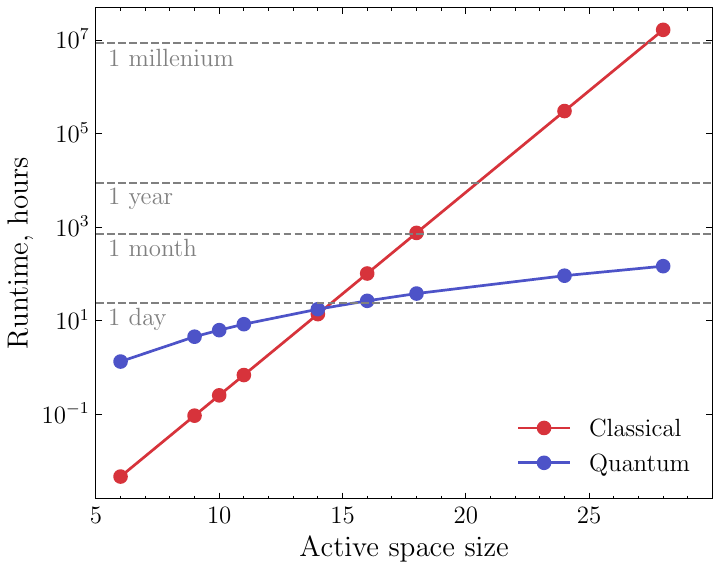}
    \caption{Comparing expected runtimes of our proposed quantum algorithm versus estimated runtime for the RAS approach for the same problem. All values are computed on a range of active spaces generated for the Li$_4$Mn$_2$O target cluster. On the quantum side we use a total qubit budget of $n_q = 350$.}
    \label{fig:classical-v-quantum}
\end{figure}
    
\textbf{Classical runtime}: To very roughly estimate the classical runtimes for the same set of systems using the RAS method requires estimating a) the number of iterations needed to converge a single eigenstate, b) the number of states required to span the XAS active region, and c) the cost per iteration. 

To estimate the cost per iteration, we can take the main cost of RAS to scale similarly to that of CAS.  There are a few CASCI scalings that can be relevant depending on system details: for concreteness we take the cost scaling to be the standard Davidson iteration scaling $O(N_{\text{det}})$, where $N_{\text{det}}$ is the number of configuration determinants in the space of interest. Considering the case of half-filling, there will be $N_{\text{det}} = \binom{N}{N/2} \sim 2^N$ determinants in the CI space, so in terms of the number of orbitals we obtain $O(2^{N})$. At the same time, very optimistically we will assume that (a) and (b) are constant prefactors that only weakly depend on system size.

The largest known configuration interaction calculation ever took 113 hours to converge one (ground) eigenstate with $N = 23$ spatial orbitals. This was done on the AI Bridging Infrastructure (ABCI) supercomputer at the University of Tokyo, the largest supercomputer in Japan and in the top 10 worldwide only a few years ago: the researchers used 256 compute nodes (roughly 10\% of the capability of ABCI at the time), each equipped with two Intel Xeon Gold 6148 CPUs, where each CPU hosts 40 logical cores and has access to 384 GB of RAM memory and a 1.6 TB solid-state drive for disk memory, all while employing an advanced parallelized implementation to tame the significant inter-node communication overheads \cite{gao2024distributed}.
Suppose an iterative Davidson-style calculation takes $M$ steps to converge. The cost of each step scales like $O(2^N)$, as argued above.
Making a generous assumption compared to what is typically seen in practice, we will assume $M$ does not scale with system size $N$.
Using the exponential scaling of the cost of each iteration and assuming constant iterations to convergence, we can scale the 113 hours runtime cited above into our case of interest with $N = 18$: we get $113 \cdot \exp(18) / \exp(23) \sim 0.76$ hours to converge the ground state with 18 orbitals, with that same device configuration.
For XAS even in much smaller active spaces (7-10 orbitals), we often need to converge hundreds or thousands of states in the XAS spectrum region \cite{delcey2019efficient}; for concreteness, we assume 1000 states are required.
We will assume that the time to simultaneously converge 1000 eigenstates in an iterative procedure can be estimated as the time to find 1 eigenstate times 1000.
Finally, we make the assumption that CVS, RAS and a projection-like approach is employed to eliminate intermediate states, so that the solver is directly computing the (core-excited) states of interest.
Combining all of this, we get a classical runtime = $0.76 \cdot 1000$ hours $\sim$ 32 days. This is likely an optimistic estimate on the classical side: a more reasonable one, once we relax some of the assumptions above, might be closer to 100 days. 

Taking both of the estimates above, we plot the classical and quantum algorithm costs for the Li$_4$Mn$_2$O cluster problem in \cref{fig:classical-v-quantum}. While these estimates have high uncertainties, by comparing a highly optimistic classical runtime estimate with a reasonable quantum clock rate we are still able to get insight into the relative scaling and comparative timing of the approaches. The current estimates suggest that the quantum algorithm will perform about a factor of $\times 30$ faster than the RAS approach for the target CAS(22e,18o) active space. This gap only widens for larger system sizes due to the significant difference in scaling between the two approaches. The classical RAS algorithm necessarily scales as $O(2^N)$ -- exponential in terms of the number of spatial orbitals -- being driven by the size of the full CI matrix, as we argued above. Meanwhile the quantum algorithm scaling is low-order polynomial, instead being chiefly determined by the number of terms in the Hamiltonian. Combining \cref{eq:global-cost,eq:perstep-cost,eq:unitary-cost,eq:zrot-cost}, the total number of gates (and hence the runtime) can be seen to scale roughly as $O(S L N^2)$: recalling that the number of fragments is chosen as $L \sim N$, and that the number of shots is roughly independent of system size, we arrive at the expected quantum scaling of $O(N^3)$ with system size, significantly more favourable than the exponential scaling of RAS.

The scaling analysis presented above implies an improvement in simulation capability of the quantum algorithm is robust: however, the exact runtime cross-over point may change for two reasons. First, it may be affected by uncertainty in the constant prefactors we estimated, either on the classical side -- for example, due to improved classical hardware or better parallelization schemes being available; or on the quantum side -- for example, due to the complexity underlying the simplified assumption of a single logical clock rate. Second, we remark that approaches other than RAS, for example those based on density matrix renormalization group (DMRG) or selective CI methods, might not scale exponentially in runtime. Due to the heuristic nature of those approaches, it is difficult to estimate their scalings without an in-depth numerical study, which is outside the scope of the current work. However, the literature suggests that computing the X-ray absorption spectrum with those methods is still a challenging task at present. For example, using DMRG the largest attempted XAS calculations include around 8 orbitals \cite{lee2023ab}.
    
\section{Conclusion\label{sec:conclusions}}

Quantum simulation of XAS can be leveraged to deduce atomic oxidation states from experimental measurements. This information is key for understanding structural degradation mechanisms
holding back the use of Li-excess materials as high-capacity battery cathodes. 
Focusing on the time-domain algorithm previously proposed for targeting XAS, here we developed a highly efficient algorithm implementation, and applied it to the industrially relevant Li$_4$Mn$_2$O cluster CAS(22e,18o) model system. Combining the use of the CDF-based product formula time evolution, the more tightly estimated number of required Trotter steps derived from a perturbation theory analysis of spectroscopy error requirements, and sampling distribution tuning, as well as several other useful optimizations, we have driven down the cost of such a calculation to only around 100 logical qubits and gate cost of $3.54\times10^8$ Toffoli-gates. This drastic cost reduction makes XAS simulation a realistic and enticing application for early fault-tolerant quantum computers. 

Going beyond the current implementation, we expect that further cost reductions are possible, both through the use of more advanced product formulas, as well as optimizing at the compilation level, both of which are clear next steps for this work. Moreover, it would be desirable to have methods that more reliably and tightly estimate the different sources of error, particularly the Trotter error, for systems outside the reach of simulators. 

Overall, our work demonstrates that XAS simulation is a highly practical yet industrially relevant use of quantum computing, and paves the way for making quantum algorithms truly useful and widely adopted in the battery industry, potentially leading to better energy storage technologies in the future.

\section{Acknowledgments}

This research used resources of the National Energy Research Scientific Computing Center (NERSC), a Department of Energy User Facility using NERSC award DDR QIS-ERCAP ERCAP0032729.

\section{Author contribution statements} 

Stepan Fomichev was the main person leading the work on the application side of the research, whereas Pablo led the work on the quantum algorithm. Both undertook most of the work. Jay Soni and Utkarsh Azad contributed software for the correctness of the simulation and resource estimation numbers. Alexander Kunitsa found errors in previous versions of the derivations in the double measurement trick appendix and provided corrected derivations. Arne-Christian and Jonathan contributed to the application and provided insightful discussions to parts of the work. Juan Miguel also contributed via discussions and coordinated the whole work.

\bibliographystyle{quantum}
\bibliography{main}
\newpage
\onecolumngrid

\appendix

\section{Compressed Double Factorization\label{app:CDF}}
Here we present the mathematical description of the Compressed Double Factorization splitting of the Hamiltonian, in more detail than is available in~\cite{cohn2021quantum}.

We start with the electronic Hamiltonian on $N$ orbitals in chemist notation:
\begin{align}
    \hat{H} = E + \sum_{p,q = 1}^N \sum_{\gamma \in \{\uparrow,\downarrow\}} (p|\kappa|q) a_{p\gamma}^\dagger a_{q\gamma}+ \frac{1}{2}\sum_{p,q,r,s =1}^N\sum_{\gamma,\tau \in \{\uparrow,\downarrow\}}(pq|rs) a_{p\gamma}^\dagger a_{q\gamma} a_{r\tau}^\dagger a_{s\tau}.
\end{align}
We numerically find $N\times N$ matrices $Z^{(\ell)}$ and $U^{(\ell)}$
\begin{align}\label{eq:(pq|rs)_app}
   (pq|rs) \approx \sum_{\ell=1}^L \sum_{k,l = 1}^N U^{(\ell)}_{pk} U^{(\ell)}_{qk} Z^{(\ell)}_{kl} U^{(\ell)}_{rl} U^{(\ell)}_{sl}
\end{align}
$Z^{(\ell)}$ is symmetric, and $U^{(\ell)}$ is orthogonal.
There is some evidence that to achieve constant error it is sufficient to fix $L = \tilde{O}(N)$~\cite{lee2021even}.
We can write the second order term as
\begin{align}
\frac{1}{2}\sum_{p,q,r,s =1}^N \sum_{\gamma,\tau \in \{\uparrow,\downarrow\}}(pq|rs) a_{p\gamma}^\dagger a_{q\gamma} a_{r,\tau}^\dagger a_{s\tau} 
   = \frac{1}{2}\sum_{\ell,kl} Z^{(\ell)}_{kl} \left( \sum_{pq,\gamma} U^{(\ell)}_{kp} a_{p\gamma}^\dagger a_{q\gamma} U^{(\ell)}_{kq} \right)\left( \sum_{rs,\tau} U^{(\ell)}_{lr} a_{r,\tau}^\dagger a_{s\tau} U^{(\ell)}_{ls} \right).
\end{align}
This factorization suggests we implement a basis rotation $\bm{U}^{(\ell)}$ dictated by $U^{(\ell)}$ to diagonalize the term in parenthesis
\begin{align}
\frac{1}{2}\sum_{p,q,r,s =1}^N \sum_{\gamma,\tau \in \{\uparrow,\downarrow\}}(pq|rs) a_{p\gamma}^\dagger a_{q\gamma} a_{r,\tau}^\dagger a_{s\tau}  
    &= \frac{1}{2}\sum_{\ell,kl,\gamma\tau} Z^{(\ell)}_{kl}  n^{(\ell)}_{k,\gamma} n^{(\ell)}_{l,\tau}\\
  &= \frac{1}{2}\sum_{\ell}  \sum_{\gamma\tau}\bm{U}^{(\ell)} \left[\sum_{kl} Z^{(\ell)}_{kl} n_{k,\gamma} n_{l,\tau} \right](\bm{U}^{(\ell)})^T.
\end{align}
Here $n^{(\ell)}_{k,\gamma}$ denotes the particle number operator in the rotated basis, and $\bm{U}^{(\ell)} := (\bm{U}^{(\ell)}_{\uparrow}\otimes \bm{1})(\bm{1}\otimes\bm{U}^{(\ell)}_{\downarrow})$ is the operator that implements the basis rotation dictated by $\bm{U}$~\cite{kivlichan2018quantum},
\begin{equation}
    \bm{U}^{(\ell)}_\gamma = \exp\left(\sum_{p,q}[\log U^{(\ell)}]_{pq} (a_{p\gamma}^\dagger a_{q\gamma}-a_{q\gamma}^\dagger a_{p\gamma})\right).
\end{equation}

Now, we substitute
\begin{equation}
    n_{k,\gamma} = \frac{\bm{1}-\sigma_{z,k\gamma}}{2}\Leftrightarrow \sigma_{z,k\gamma} = \bm{1} - 2n_{k,\gamma},
\end{equation}
where $\sigma_{z,k\gamma}$ is the Pauli $Z$ operator acting on orbital $k,\gamma$;
obtaining
\begin{align}\label{eq:part_energy_shift}
&\frac{1}{2}\sum_\ell\sum_{kl,\gamma\tau} \bm{U}^{(\ell)} \left(\sum_{kl} Z^{(\ell)}_{kl} n_{k,\gamma} n_{l,\tau}\right) (\bm{U}^{(\ell)})^T\\
     = &\frac{1}{2}\left(\sum_\ell \sum_{kl}Z_{lk}^{(\ell)}\right)\bm{1}\\
     \label{eq:one_body_correction}
     - &\frac{1}{2}\sum_\ell \sum_{k,\gamma} \bm{U}^{(\ell)} \left(\sum_{l} Z^{(\ell)}_{kl}\right) \sigma_{z,k\gamma} (\bm{U}^{(\ell)})^T\\
     + &\frac{1}{8} \sum_\ell\sum_{\gamma\tau} \bm{U}^{(\ell)} \left(\sum_{kl} Z^{(\ell)}_{kl} \sigma_{z,k\gamma} \sigma_{z,l\tau}\right) (\bm{U}^{(\ell)})^T
\end{align}
We may take the third term,
\begin{align}
    &\frac{1}{8} \sum_\ell\sum_{\gamma\tau} \bm{U}^{(\ell)} \left(\sum_{kl} Z^{(\ell)}_{kl} \sigma_{z,k\gamma} \sigma_{z,l\tau}\right) (\bm{U}^{(\ell)})^T\\
    \label{eq:two_body_term}
    = & \frac{1}{8} \sum_\ell\sum_{(k,\gamma)\neq (l,\tau)} Z^{(\ell)}_{kl} \bm{U}^{(\ell)}   \sigma_{z,k\gamma} \sigma_{z,l\tau} (\bm{U}^{(\ell)})^T\\
    + &\frac{1}{4} \sum_\ell\sum_{k} Z^{(\ell)}_{kk} \bm{1}
\end{align}
because $\sigma_{z,k\gamma}^2 = \bm{1}$.
\cref{eq:two_body_term} represents the two-body term we will actually implement. 

The factor of $1/2$ in front of~\cref{eq:one_body_correction} is due to the two factors of $1/2$ in $n_{k,\gamma}$ and $n_{l,\tau}$ being cancelled by the two products in $n_{k,\gamma}n_{l,\tau}$ that result in $\sigma_{z,k\gamma}$ (or $\sigma_{z,l\tau}$) as well as the sum over $\tau$ (or $\gamma$). \cref{eq:one_body_correction} will be merged with the one-body term. To do so we will return it to its $n_{k,\gamma}$ form 
\begin{align}
    &- \frac{1}{2}\sum_\ell \sum_{k,\gamma} \bm{U}^{(\ell)}_{\gamma} \left(\sum_{l} Z^{(\ell)}_{kl}\right) \sigma_{z,k\gamma} (\bm{U}^{(\ell)}_{\gamma})^T\\
    = &- \frac{1}{2}\sum_\ell \sum_{k,\gamma} \bm{U}^{(\ell)}_{\gamma} \left(\sum_{l} Z^{(\ell)}_{kl}\right) (\bm{1}-2n_{k,\gamma} )(\bm{U}^{(\ell)}_{\gamma})^T\\
    = &-\sum_\ell \sum_{kl} Z_{kl}^{(\ell)} \bm{1}+\sum_\ell \sum_{k,\gamma}  \left(\sum_{l} Z^{(\ell)}_{kl}\right)\bm{U}^{(\ell)}_{\gamma}  n_{k,\gamma}(\bm{U}^{(\ell)}_{\gamma})^T
\end{align}
Combining the last two lines with~\eqref{eq:part_energy_shift}, we obtain an overall energy shift
\begin{align}
    -\frac{1}{2}\sum_\ell \sum_{kl} Z_{kl}^{(\ell)} \bm{1}
\end{align}
and a one-body correction
\begin{align}
    \sum_\ell \sum_{k,\gamma}  \left(\sum_{l} Z^{(\ell)}_{kl}\right)  n^{(\ell)}_{k,\gamma} 
    = \sum_\ell \sum_{k}  \left(\sum_{l} Z^{(\ell)}_{kl}\right)  \sum_{pq,\gamma} U^{(\ell)}_{pk} U^{(\ell)}_{qk}a_{p\gamma}^\dagger a_{q\gamma}
\end{align}
to the one-body term. We next have to diagonalize the joint one-body operator
\begin{align}
    T &= \sum_{pq,\gamma} t_{pq}a_{p\gamma}^\dagger a_{q\gamma}
\end{align}
with
\begin{align}\label{eq:t_pq}
    t_{pq} = (p|\kappa|q) + \sum_\ell \sum_{k}  \left(\sum_{l} Z^{(\ell)}_{kl}\right) U^{(\ell)}_{pk} U^{(\ell)}_{qk}
\end{align}
then, we write
\begin{align}\label{eq:t_pq_UZ0}
    t_{pq} = \sum_k U^{(0)}_{pk} Z^{(0)}_{k} U^{(0)}_{qk}
\end{align}
which backsubstituting gives us
\begin{align}
    T &= \sum_{pq,\gamma} t_{pq}a_{p\gamma}^\dagger a_{q\gamma}\\
    &= \sum_{pq,\gamma} \sum_k  Z^{(0)}_{k} (U^{(0)})^T_{kp} a_{p\gamma}^\dagger a_{q\gamma} U^{(0)}_{qk} \\
    &= \sum_k Z^{(0)}_{k} \sum_\gamma n^{(0)}_{k,\gamma}
\end{align}
Again, this means that in the basis rotation dictated by $U^{(0)}$ the one-body operator becomes diagonal, so we implement it as
\begin{align}
    T &=  \sum_k Z^{(0)}_{k} \sum_\gamma \bm{U}^{(0)} n_{k,\gamma} (\bm{U}^{(0)})^T\\
    &=\sum_k Z^{(0)}_{k} \sum_\gamma \bm{U}^{(0)} \frac{\bm{1}-\sigma_{z,k\gamma}}{2} (\bm{U}^{(0)})^T\\
    &=\sum_k Z^{(0)}_{k}\bm{1} -\frac{1}{2}\sum_k Z^{(0)}_{k} \sum_\gamma \bm{U}^{(0)} \sigma_{z,k\gamma} (\bm{U}^{(0)})^T
\end{align}
The first term will be another global phase, and the last one corresponds to the one-body term to be implemented. Overall, we implement the Hamiltonian
\begin{align}
    \hat{H}  &= \left(E + \sum_k Z_k^{(0)} -\frac{1}{2}\sum_{\ell,kl}Z_{kl}^{(\ell)}+\frac{1}{4} \sum_{\ell,k} Z^{(\ell)}_{kk} \right)\bm{1}\nonumber\\ \label{eq:CDF_Hamiltonian}
    &-\frac{1}{2}\bm{U}^{(0)} \left[\sum_k Z^{(0)}_{k} \sum_\gamma  \sigma_{z,k\gamma} \right](\bm{U}^{(0)})^T\\
    &+\frac{1}{8}\sum_\ell  \bm{U}^{(\ell)} \left[\sum_{(k,\gamma)\neq(l,\tau)}  \left( Z^{(\ell)}_{kl} \sigma_{z,k\gamma} \sigma_{z,l\tau}\right) \right](\bm{U}^{(\ell)})^T.\nonumber
\end{align}
where the matrices $U^{(\ell)}$ and $Z^{(\ell)}$ are defined in~\cref{eq:(pq|rs)_app} and~\cref{eq:t_pq_UZ0}. Note that the definition of $U^{(0)}$ and $Z^{(0)}$ differ from the main text because here we also include the one-body contributions from the 2-particle integrals,~\cref{eq:t_pq}.

\begin{figure*}
\begin{minipage}{0.45\textwidth}
\includegraphics[width=\textwidth]{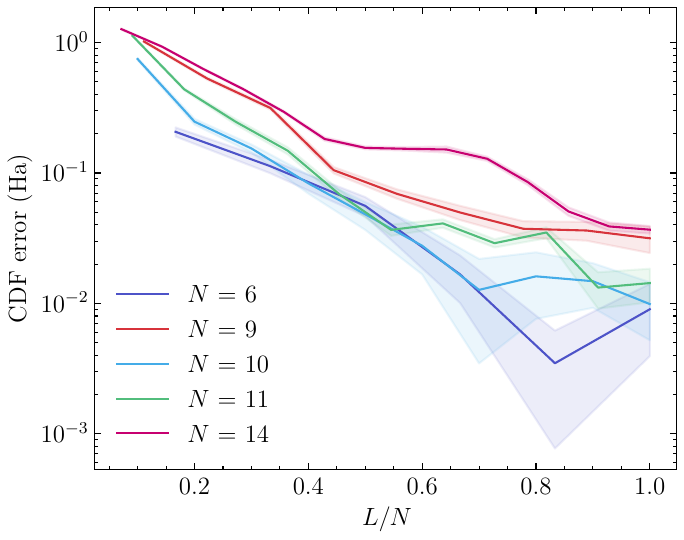}
    \caption{Mean and standard deviation of the error of the CDF approximation of the LiMnO Hamiltonian over the first 25 eigenvalues, as a function of the rank considered $L$ and the number of orbitals in the active space $N$. Selecting $L = N$ is thus approximately sufficient to satisfy the target precision of 1 eV $\approx$ 0.037 Ha.}
    \label{fig:CDF-error-LiMnO}
\end{minipage}
\hspace{5pt}
\begin{minipage}{0.45\textwidth}
    \centering
\includegraphics[width=\linewidth]{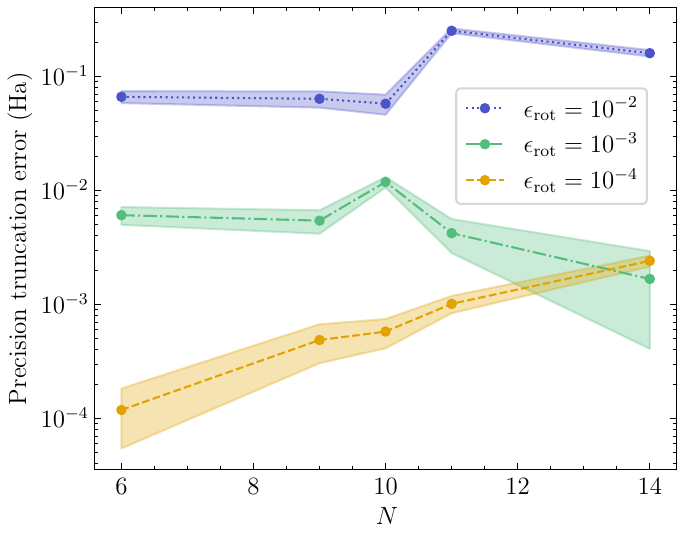}
    \caption{Mean and standard deviation of the error over the lowest 25 eigenvalues of the LiMnO CDF Hamiltonian with $L = N$, due to restricting the precision of the CDF matrices. From these results we expect $\epsilon_{\text{rot}} = 10^{-3}$ to be sufficient to achieve $1$ eV $\approx$ $0.037$ Ha.}
    \label{fig:rotation-error-LiMnO}
\end{minipage}
\end{figure*}

The key free parameter to fix in the CDF decomposition of the Hamiltonian is the number of fragments $L$. In \cref{fig:CDF-error-LiMnO}, we numerically demonstrate that the choice of $L \sim N$ where $N$ is the number of spatial orbitals of the system is usually sufficient to keep errors manageable. Focusing on the model cluster LiMnO, we do this by producing different CDF fragmentations with varying number of fragments $L$: for each of these, we then contract it back to the corresponding two-electron integrals $(pq|rs)$, then use standard full configuration interaction methods to calculate the lowest 25 eigenvalues and compare them to those of the exact Hamiltonian. As can be seen from the figure, keeping $L \approx N$ is sufficient to keep the errors below our application-relevant error threshold of around 1 eV.

\begin{figure}[t]
    \centering
    \includegraphics[width=0.5\linewidth]{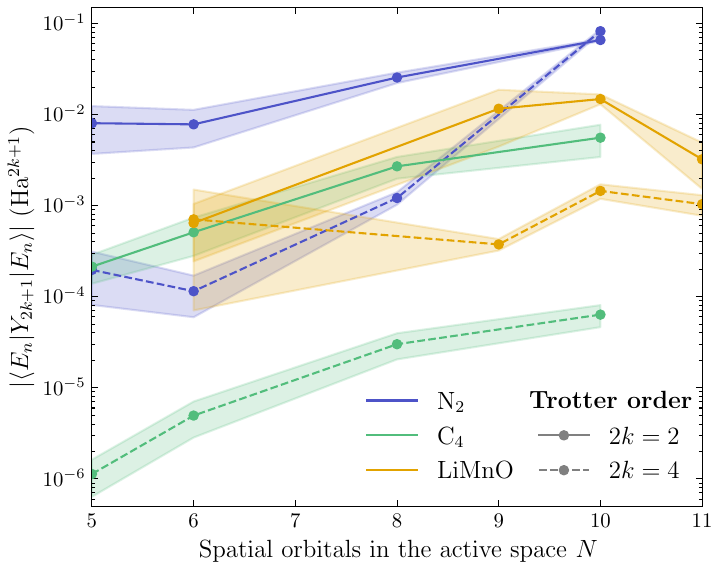}
    \caption{Error of the second and fourth order formulas in the first five eigenvalues of CDF Hamiltonians for active spaces of N$_2$, C$_4$ and LiMnO. $N$ represents the number of spatial orbitals of the Hamiltonian. Based on these results, the resource estimates in~\cref{tab:resources} assume $|\braket{Y_3}|<$ 1 Ha$^3$. The standard deviation of fourth order formula for the $N = 6$ LiMnO cluster is larger than the average, so it is truncated to a tenth of the mean.}
    \label{fig:trotter_error}
\end{figure}

\section{Error analysis\label{app:error}}

\subsection{Trotter error}\label{appssec:Trotter_error}

In the sections above we have considered that we want to simulate each $j$ samples with a constant amount of error (or increasing with the sampling error). Here we make the case for taking a constant number of Trotter steps with each $j$. In other words, we are not simulating $e^{-ij\tau H} + \epsilon $ for a small constant error $\epsilon$. Instead, we are simulating
\begin{equation}
    U_2^j(\tau) = e^{-i\tau j \hat{H}  + (-i \tau)^{2k+1} j Y_{2k+1} + (-i \tau)^{2k+3} j Y_{2k+3} +\ldots }
\end{equation}
We are assuming we take the Trotter step $\Delta:= \tau$ for simplicity, but the generalization is straightforward.
By using the algorithm, we are approximating
\begin{align}
    - \eta \operatorname{Im} G_\rho(\omega, H) &= \frac{\eta \tau}{2\pi}\sum_{j=-\infty}^{+\infty} e^{-|j|\tau \eta}\braket{m_\rho| e^{-ij\tau H}|m_\rho} e^{+ij\tau \omega}\\
    &= \frac{\eta \tau}{2\pi}\sum_{j=-\infty}^{+\infty} e^{-|j|\tau \eta}\braket{m_\rho| e^{-ij\tau \sum_l E_l^{(0)} \ket{E_l^{(0)}}\bra{E_l^{(0)}}}|m_\rho} e^{+ij\tau \omega}\\
    &= \frac{\eta \tau}{\pi} \sum_{l} \frac{\eta}{\eta^2 + (E^{(0)}_l-\omega)^2} |\braket{m_\rho|E^{(0)}_l}|^2
\end{align}
where $\ket{E^{(0)}_l}$ are eigenstates of $H$ with eigenvalues $E^{(0)}_l$. Using (non-degenerate) perturbation theory to first order we approximate $\hat{H}'  = \hat{H}  + (-i \tau)^{2k} Y_{2k+1} + \ldots$
\begin{align}
    - \eta \operatorname{Im} G_\rho(\omega, \hat{H}' ) = \frac{\eta \tau}{2\pi}\sum_{j=-j_{\max}}^{+j_{\max}} e^{-|j|\tau \eta}\braket{m_\rho|U_2^j(\tau)|m_\rho} e^{+ij\tau \omega} \approx \frac{\eta \tau}{\pi} \sum_l \frac{\eta}{\eta^2 + (E_l-\omega)^2} |\braket{m_\rho|E_l}|^2,
\end{align}
where
\begin{align}
    \ket{E_l} &= \ket{E^{(0)}_l} - \tau^{2k} \sum_{k\neq l} \frac{\braket{E_k^{(0)}|Y_{2k+1}|E_l^{(0)}}}{E_k^{(0)}-E_l^{(0)}}
    \ket{E_k^{(0)}}+O(\tau^{2k+2})\\
    E_l &= E_l^{(0)} - \tau^{2k}\braket{E_l^{(0)}|Y_{2k+1}|E_l^{(0)}} + O(\tau^{2k+2})
\end{align}
This last equation is true even if states are degenerate. If eigenstates are degenerate (but the degeneracy is lifted by $Y_{2k+1}$) then~\cite[Eq. 1.2.41]{Zwiebach18Perturbation}
\begin{align}
    \ket{E_{l,m}} = \ket{E^{(0)}_{l,m}}& - \tau^{2k} \sum_{k\neq l} \frac{\braket{E_k^{(0)}|Y_{2k+1}|E_{l,m}^{(0)}}}{E_k^{(0)}-E_{l}^{(0)}}
    \ket{E_k^{(0)}}\\
    &-\tau^{2k} \sum_{n\neq m}\frac{\ket{E_{l,n}^{(0)}}}{E^{(1)}_{l,m}-E^{(1)}_{l,n}}\sum_{k\neq l}\frac{\braket{E_{l,m}^{(0)}|Y_{2k+1}|E_k^{(0)}}\braket{
    E_{k}^{(0)}|Y_{2k+1}|E_{l,n}^{(0)}}}{E_k^{(0)}-E_{l}^{(0)}}+O(\tau^{2k+2})
\end{align}
where $\ket{E_{l,m}^{(0)}}$ are chosen orthogonal in each degenerate subspace, which is indexed by $l$. The error induced by Trotter is
\begin{equation}
    \epsilon_{\text{Tr}} = |C_{\eta}(\hat{H}' , \omega) - C_{\eta}(H, \omega)|
\end{equation}
Let 
\begin{align}
\epsilon_l = E_l - E_l^{(0)} = \tau^{2k} |\braket{E_l^{(0)}|Y_{2k+1}|E_l^{(0)}}| + O(\tau^{2k+2})\\
\varepsilon_l = \braket{m_\rho|E_l} - \braket{m_\rho|E^{(0)}_l} = - \tau^{2k} \sum_{k\neq l} \frac{\braket{E_k^{(0)}|Y_{2k+1}|E_l^{(0)}}}{E_k^{(0)}-E_l^{(0)}}
    \braket{m_\rho|E_k^{(0)}}+O(\tau^{2k+2}).
\end{align}
where in the second equation we used the non-degenerate equations but a similar procedure could be used for the degenerate case. Then, we have
\begin{align}
    - \eta \operatorname{Im} G_\rho(\omega, \hat{H}' ) &= \frac{\eta \tau}{2\pi}\sum_{j=-j_{\max}}^{+j_{\max}} e^{-|j|\tau \eta}\braket{m_\rho|U_2^j(\tau)|m_\rho} e^{+ij\tau \omega} \nonumber\\
    &\approx \frac{\eta \tau}{\pi} \sum_l \frac{\eta}{\eta^2 + (E_l-\omega)^2} |\braket{m_\rho|E_l}|^2.\\
    &=  \frac{\eta \tau}{\pi} \sum_l \frac{\eta}{\eta^2 + (E_l^{(0)}-\omega + \epsilon_l)^2} |\braket{m_\rho|E_l^{(0)}}+ \varepsilon_l|^2.
\end{align}

These expressions are independent of the $j_{\max}$ considered to approximate the dampened Fourier transform. As such, independently of $j_{\max}$, there is a value of $\tau$ that makes the result sufficiently accurate, and allows us to take a constant number of Trotter steps per $j$. Specifically, if we want the error in the eigenvalues to be smaller than $\epsilon_l = 0.05$ Ha, it is sufficient to take $\tau \approx \max_{l}\left|\frac{|\braket{E_l|Y_{2k+1}|E_l}|}{\epsilon_l}\right|^{1/2k}$.

\subsection{Measurement noise} \label{appssec:measurement_noise}
Now, we want to understand what is the effect of the sampling error in the result. Recall that the variance of a binomial distribution is given by
\begin{equation}
    \sigma^2 = \frac{p(1-p)}{N}\leq \frac{1}{4N}
\end{equation}
The noise in the signal can be characterized by measuring the real and imaginary components of $\braket{m_\rho|U_2(\tau)^j|m_\rho}$, via Hadamard tests that can be converted into such value by $p(0)-p(1) = 2p(0)-1$ where $p(0)$ is the probability of measuring $0$ and similarly for $1$.

A key component of our analysis is that since the Fourier transform is linear, we can compute the Fourier transform of the signal and noise separately and add the result. For the noise, we are computing
\begin{equation}
    -\eta  \operatorname{Im} G_\rho(\omega, \text{noise}) = \frac{\tau\eta}{2\pi} \sum_j e^{-|j|\tau \eta} e^{-i j \tau \omega} (X_j+iY_j)
\end{equation}
where $X_j$ and $Y_j$ are binomial distributions with expected value 0 and variance $\sigma^2 = \frac{\sum_k e^{-|k|\tau \eta}}{S e^{-|j|\tau \eta}}$ because we take the number of shots at a given $j$ to be $S_j = S \frac{e^{-|j|\tau \eta}}{\sum_k e^{-|k|\tau \eta}}$. The factor of 4 we saw above is cancelled by the multiplication by $2$ in the probability of measuring $0$, as well as adding the variance of the real and imaginary components. Computing the result, the variance of the sum in the discrete Fourier transform is the sum of variances. That is
\begin{align}
    \epsilon_{\text{meas}}^2 &= \left(\frac{\eta\tau}{2\pi}\right)^2\sum_j |e^{-i j \tau \omega}|^2\left(\frac{\sigma^2\{e^{-|j|\tau \eta} X_j\}}{S_j}+\frac{\sigma^2\{e^{-|j|\tau \eta} Y_j\}}{S_j}\right) \nonumber\\
    &\leq \left(\frac{\eta\tau}{2\pi}\right)^2\sum_j e^{-2|j|\tau \eta} |e^{-i j \tau \omega}|\left(\frac{\sigma^2\{X_j\}}{S_j}+\frac{\sigma^2\{Y_j\}}{S_j}\right)\\
    &\leq \left(\frac{\eta\tau}{2\pi}\right)^2\sum_j e^{-2|j|\tau \eta} \frac{\sum_k e^{-|k|\tau\eta}}{S e^{-|j|\tau\eta}} \Rightarrow S = \left(\frac{\eta\tau A}{2\pi\epsilon_{\text{meas}}}\right)^2,
\end{align}
where $A = \sum_{k=0}^{j_{\max}} e^{-|k|\tau \eta}$, consistent with the notation in the main text.

If we had allocated the same sample budget to each $j$, we would have gotten instead
\begin{align}
    \epsilon_{\text{meas}}^2 &= \left(\frac{\eta\tau}{2\pi}\right)^2\sum_j |e^{-i j \tau \omega}|^2\left(\frac{\sigma^2\{e^{-|j|\tau \eta} X_j\}}{S/j_{\max}}+\frac{\sigma^2\{e^{-|j|\tau \eta} Y_j\}}{S/j_{\max}}\right) \nonumber\\
    &\leq  \left(\frac{\eta\tau}{2\pi}\right)^2\sum_j e^{-2|j|\tau \eta} \frac{j_{\max}}{S}\left(\sigma^2\{X_j\}+\sigma^2\{ Y_j\}\right)\\
    &\leq \left(\frac{\eta\tau}{2\pi}\right)^2 \sum_j e^{-2|j|\tau \eta}\frac{j_{\max}}{2S} \Rightarrow S = \frac{j_{\max}}{2} \sum_j e^{-2 j\eta \tau}\left(\frac{\eta\tau}{2\pi\epsilon_{\text{meas}}}\right)^2.
\end{align}

\subsubsection{Optimizing the sampling cost}
Let us now find the optimal sampling distribution. We know the cost of one shot for $G(\tau j)$ is some fixed cost for state preparation $C_S$ and the cost of evolving for $j\frac{\tau}{\Delta}$ steps, where each step costs $ C_{\text{Trot}}$. That is, the cost of one sample of $G(\tau j)$ is
\begin{align}
    C_j = C_S + \frac{j\tau}{\eta}C_{\text{Trot}}
\end{align}

Let us have a sampling distribution
\begin{align}
    S_j = S\frac{e^{-\alpha |j|\tau\eta}}{\sum_k e^{-\alpha |k|\tau\eta}}.
\end{align}
where $\alpha$ is a parameter to be optimized. The number of shots needed under this distribution to achieve error $\epsilon_{\text{meas}}$ is
\begin{align}
    \epsilon_{\text{meas}}^2 \leq \left(\frac{\eta\tau}{2\pi}\right)^2 \sum_j e^{-2|j|\tau \eta} \frac{\sum_k e^{-\alpha|k|\tau\eta}}{S e^{-\alpha|j|\tau\eta}} \Rightarrow S &= \left(\frac{\eta\tau}{2\pi\epsilon_{\text{meas}}}\right)^2 \sum_j e^{(\alpha-2)|j|\tau\eta}\sum_k e^{-\alpha|k|\tau\eta}\\
    \Rightarrow S_j &= \left(\frac{\eta\tau}{2\pi\epsilon_{\text{meas}}}\right)^2 e^{-\alpha|j|\tau\eta} \sum_k e^{(\alpha-2)|k|\tau\eta}.
\end{align}
The cost over all shots is
\begin{align}
    C &= \sum_j S_j C_j = \left(\frac{\eta\tau}{2\pi}\right)^2\sum_j\left(C_S + \frac{j\tau}{\eta}C_{\text{Trot}}\right)\frac{e^{-\alpha|j|\tau\eta} \sum_k e^{(\alpha-2)|k|\tau\eta} }{\epsilon_{\text{meas}}^2} \nonumber\\
    &= \left(\frac{\eta\tau}{2\pi}\right)^2 \sum_{j,k}\left(C_S + \frac{j\tau}{\eta}C_{\text{Trot}}\right)\frac{e^{-\alpha(|j|-|k|)\tau\eta}  e^{-2|k|\tau\eta} }{\epsilon_{\text{meas}}^2}.
\end{align}
For the parameters in the problem, eg $j_{\max} = 200$, $\eta = 0.05$ Ha, $\tau = \frac{\pi}{4}$ Ha$^{-1}$, $D = 10^4$ and the LiMnO cluster, the minimum is achieved for $\alpha = 1.338$ saving a multiplicative factor of $\times 1.178$ speedup at constant $\epsilon_{\text{meas}}$ vs taking $\alpha = 1$, and a factor of $\times 17.9$ with respect to the uniform distribution, e.g. $\alpha = 0$. Note that this increases the number of shots, but reduces the total cost.

\subsection{Modelling discrete-time Fourier transform error \label{appssec:truncation_error}}

There are two more sources of error. The first one is related to truncating the Fourier transform integral to the interval $\tau j\in(-\tau  j_{\max}, +\tau j_{\max})$ instead of $\tau j \in(-\infty, +\infty)$.
The error can be bounded by considering the norm of all that we left out:
\begin{align}\label{eq:Truncation_error}
    \epsilon_{\text{trunc}} &\leq \frac{2\eta \tau}{2\pi} \int_{j_{\max}}^\infty e^{-|j|\tau \eta}\cdot|e^{-ij\tau w}|\cdot\|\braket{m_{\rho}|e^{-i\tau j H}|m_{\rho}}\| dj \\
    &\leq \frac{\eta \tau}{\pi} \int_{j_{\max}}^\infty e^{-|j|\tau \eta} dj = \frac{\eta \tau}{\pi\tau \eta}\int_{j_{\max}\tau \eta}^\infty e^{-|k|} d k = \frac{e^{-|j_{\max}\tau\eta|}}{\pi},
\end{align}
where we used the change of variables $k = j\eta \tau$. As such,
\begin{align}
    j_{\max} \geq \frac{-\log (\epsilon_{\text{trunc}} \pi )}{\tau \eta} = \tilde{O}((\tau \eta)^{-1}).
\end{align}

Additionally, we can bound the error from the time discretization with the error of a Riemann sum that approximates the Fourier transform integral. From~\cref{eq:Riemann_sum_error} we get
\begin{align}
    \epsilon_{\text{disc}} &\leq \frac{(j_{\max}\tau)^3}{24\pi j_{\max}^2}\max_{j}\left|\frac{d^2}{dt^2}\braket{\hat{m}_\rho|e^{-i t (H-w-i\eta)}|\hat{m}_\rho}\right|_{t =j\tau}\leq \frac{j_{\max}\tau^3}{24\pi}\left|H-w-i\eta\right|^2.
\end{align}
This provides further justification to the selection $\tau = \tilde{O}(j_{\max}^{-1}) = \tilde{O}(\|H\|_\omega^{-1})$, for $\|H\|_\omega$ an energy window that prevents aliasing of the optically-active states.

\section{Double measurement trick}\label{appsec:double_measurement_trick}

\subsection{Non-unitary state preparation}

\begin{figure}
    \[
    \Qcircuit @C=1em @R=1em {
        \lstick{\ket{0}} & \qw & \qw & \qw & \qw & \gate{H} & \ctrl{3} & \gate{H} & \meter & \cw \\
        \lstick{\ket{0}} & \gate{H} & \ctrl{1} & \gate{S^\dagger} \gategroup{2}{4}{2}{4}{.7em}{--} & \gate{H} & \meter & \cw & \cw & \cw & \\
        \lstick{\ket{\rho}} & \qw & \gate{U} & \qw & \qw & \qw & \qswap & \qw & \qw & \qw \\
        \lstick{\ket{\rho}} & \qw & \qw & \qw & \qw & \qw & \qswap & \qw & \qw & \qw
    }
    \]
    \caption{Double measurement Hadamard circuit for non-unitary state preparation. It performs a $\mathtt{SWAP}$ test on the output of the Hadamard circuit, returning additional information about the expectation value $\bra{\rho} U \ket{\rho}$, when the state preparation $\hat{m}_\rho$ is not unitary.}
    \label{fig:double_measurement}
\end{figure}
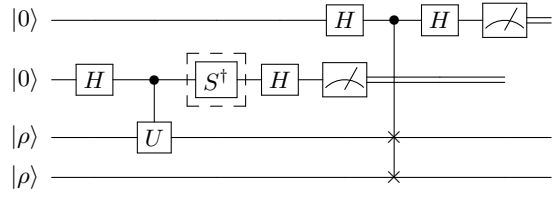

Instead of separately measuring the real and imaginary components in a Hadamard test, we can measure the cost of one almost directly from the other, as shown in~\cref{fig:double_measurement}. The trick is to recycle the output of the Hadamard test, either 
    \begin{equation}
       \ket{\psi_\Re} = \frac{(1\pm U)\ket{\rho}}{\sqrt{\braket{\rho|(1\pm U)^\dagger(1\pm U)|\rho}}} = \frac{(1\pm U)\ket{\rho}}{\sqrt{2 \pm (\braket{\rho|U|\rho}  + \braket{\rho|U^\dagger|\rho})}}  = \frac{(1\pm U)\ket{\rho}}{\sqrt{2(1 \pm \Re \braket{\rho|U|\rho})}}.
    \end{equation}
    in a swap test, where $\rho$ is the initial state and $U = e^{-i\tau j H}$. The other state involved in the swap test is the initial state $\ket{\rho}$ itself.

    The probability of measuring $0$ in the swap test is then
    \begin{equation}
        P_\pm (0) = \frac{1}{2}+\frac{1}{2}|\braket{\psi_\Re|\rho}|^2 = \frac{1}{2}+\frac{1}{2}\frac{|1\pm \braket{\rho|U|\rho}|^2}{2(1 \pm \Re \braket{\rho|U|\rho})}.
    \end{equation}
    As a consequence,
    \begin{align}
        |1\pm \braket{\rho|U|\rho}|^2 = 2(2P_\pm (0) -1)(1\pm\Re(\braket{\rho|U|\rho}))\\
        \Rightarrow|1\pm (\Re\braket{\rho|U|\rho}) + 1i\cdot (\Im\braket{\rho|U|\rho})|^2 = 2(2P_\pm (0) -1)(1\pm\Re(\braket{\rho|U|\rho}))\\
        \Rightarrow |\Im\braket{\rho|U|\rho} |= \sqrt{2(2P_\pm (0) -1)(1\pm\Re(\braket{\rho|U|\rho}))-|1\pm \Re\braket{\rho|U|\rho}|^2}
        \label{eq:abs(Im<U>)}
    \end{align}
    where the $\pm$ depends on the result of the Hadamard test, $+$ for $\ket{0}$ and $-$ for $\ket{1}$. The sign should of the $\Re\braket{\rho|U|\rho}$ can only be determined in the Hadamard test. The error of this estimate is
    \begin{align}
    \epsilon_{\Im, \pm}^2 &= \frac{16(1\pm\Re{\langle\rho|U|\rho\rangle})^2\epsilon_{S,\pm}^2 + [\pm 2(2P_{\pm}(0)-1) \mp 2(1\pm\Re{\langle\rho|U|\rho\rangle})]^2\epsilon_H^2 }{4|\Im\braket{\rho|U|\rho} |^2}\\
    &=\frac{4(1\pm\Re{\langle\rho|U|\rho\rangle})^2\epsilon_{S,\pm}^2 + (\pm 2P_{\pm}(0)-\Re{\langle\rho|U|\rho\rangle}\mp 2)^2 \epsilon_H^2 }{|\Im\braket{\rho|U|\rho} |^2}\\
    &=\frac{4(1\pm\Re{\langle\rho|U|\rho\rangle})^2\epsilon_{S,\pm}^2 + ( 2P_{\pm}(0)-2\mp\Re{\langle\rho|U|\rho\rangle})^2 \epsilon_H^2 }{|\Im\braket{\rho|U|\rho} |^2}.
    \label{eq:epsilon_Impm}
\end{align}

    If instead of measuring the real component in the swap test we are measuring the imaginary component, then the states are
    \begin{equation}
       \ket{\psi_\Im} = \frac{(1\mp i U)\ket{\rho}}{{\sqrt{\braket{\rho|(1\mp i U)^\dagger(1\mp i U)|\rho}}}} =\frac{(1\mp i U)\ket{\rho}}{{\sqrt{2 \mp i(\braket{\rho|  U|\rho} -  \braket{\rho|  U^\dagger|\rho})}}} =\frac{(1\mp i U)\ket{\rho}}{{\sqrt{2(1 \pm \Im \braket{\rho|U|\rho})}}}.
    \end{equation}
    We have
        \begin{equation}
        P_\pm (0) = \frac{1}{2}+\frac{1}{2}|\braket{\psi_\Im|\rho}|^2 = \frac{1}{2}+\frac{1}{2}\frac{|1\mp i \braket{\rho|U|\rho}|^2}{2(1 \pm \Im \braket{\rho|U|\rho})}
    \end{equation}
    and 
    \begin{align}
        |1\mp i\braket{\rho|U|\rho}|^2 = 2(2P_\pm (0) -1)(1 \pm \Im \braket{\rho|U|\rho})\\
        \Rightarrow|1\mp i (\Re\braket{\rho|U|\rho} + 1i\cdot (\Im\braket{\rho|U|\rho}|^2 = 2(2P_\pm (0) -1)(1 \pm \Im \braket{\rho|U|\rho})\\
        \Rightarrow |\Re\braket{\rho|U|\rho}| =  \sqrt{2(2P_\pm (0) -1)(1 \pm \Im \braket{\rho|U|\rho})-|1\pm \Im\braket{\rho|U|\rho}|^2}
        \label{eq:abs(Re<U>)}
    \end{align}
with the same convention for the $\pm$ signs as in the real case.
The error of this estimate is
\begin{align} \label{eq:epsilon_Repm}
    \epsilon_{\Re, \pm}^2 &=\frac{4(1\pm\Im{\langle\rho|U|\rho\rangle})^2\epsilon_{S,\pm}^2 + ( 2P_{\pm}(0)-2\mp\Im{\langle\rho|U|\rho\rangle})^2 \epsilon_H^2 }{|\Re\braket{\rho|U|\rho} |^2}.
\end{align}

The number of measurements $N_{\Re, \pm}$ (or $N_{\Im,\pm}$) taken is 
\begin{align}
    N_{\Re, \pm} \approx \frac{N}{2}(1\pm \Re{\langle\rho|U|\rho\rangle})\qquad \text{or} \qquad 
    N_{\Im,\pm} \approx  \frac{N}{2}(1\pm \Im{\langle\rho|U|\rho\rangle})
\end{align}
for~\cref{eq:epsilon_Impm} or~\cref{eq:epsilon_Repm} respectively.

Using error propagation theory, for independent measurements $\{x_i\}$ we can estimate the expected value and error as
\begin{align}
    \mathbb{E}({x}) = \frac{\sum_ix_i\sigma^{-2}_i}{\sum_j\sigma_j^{-2}},
\qquad 
    \sigma^2(x) = \frac{1}{\sum_j\sigma_j^{-2}}.
\end{align}
The error in a binomial distribution after $N$ measurements can be bounded as $\sqrt{p(1-p)/N}$. This means that $\epsilon_H^{2} \leq \frac{1}{2N}$, and $\epsilon_{S,\pm}^{2} \leq \frac{1}{2N_{\Re,\pm}} \approx \frac{2}{2N(1\pm\Re(\braket{\rho|U|\rho}))}$ in~\cref{eq:epsilon_Impm} (equivalently for the imaginary component in~\cref{eq:epsilon_Repm}). As a consequence,
\begin{align}
    \epsilon_{\Re, \pm}^2 &=\frac{4(1\pm\Im{\langle\rho|U|\rho\rangle})^2\epsilon_{S,\pm}^2 + ( 2P_{\pm}(0)-2\mp\Im{\langle\rho|U|\rho\rangle})^2 \epsilon_H^2 }{|\Re\braket{\rho|U|\rho} |^2}\\
    &\leq \frac{8(1\pm\Im{\langle\rho|U|\rho\rangle}) + ( 2P_{\pm}(0)-2\mp\Im{\langle\rho|U|\rho\rangle})^2  }{|\Re\braket{\rho|U|\rho} |^2}\frac{1}{2N}\\
    &\approx \frac{8(1\pm\Im{\langle\rho|U|\rho\rangle}) + ( \frac{|1\mp i \braket{\rho|U|\rho}|^2}{2(1 \pm \Re \braket{\rho|U|\rho})}-1 \mp\Im{\langle\rho|U|\rho\rangle})^2  }{|\Re\braket{\rho|U|\rho} |^2}\frac{1}{2N}.
\end{align}

Similar results may be achieved for the imaginary part. This implies that the number of Hadamard measurements we would require to achieve the same precision would be
\begin{align}
N_{\Re} ={}& N \Bigg[ 1
+\frac{|\Re\braket{\rho|U|\rho} |^2}{8(1+\Im{\langle\rho|U|\rho\rangle}) + ( \frac{|1-i \braket{\rho|U|\rho}|^2}{2(1 + \Re \braket{\rho|U|\rho})}-1 -\Im{\langle\rho|U|\rho\rangle})^2  }  \nonumber\\
&+ \frac{|\Re\braket{\rho|U|\rho} |^2}{8(1-\Im{\langle\rho|U|\rho\rangle}) + ( \frac{|1+i \braket{\rho|U|\rho}|^2}{2(1 - \Re \braket{\rho|U|\rho})}-1 +\Im{\langle\rho|U|\rho\rangle})^2  } \Bigg].\label{eq:N_R_doublemeasurement}
\end{align}

Similarly, for the number of equivalent measurements in of the imaginary component is
\begin{align}
N_{\Im} ={}& N \Bigg[ 1
+\frac{|\Im\braket{\rho|U|\rho} |^2}{8(1+\Re{\langle\rho|U|\rho\rangle}) + ( \frac{|1+ \braket{\rho|U|\rho}|^2}{2(1 + \Im \braket{\rho|U|\rho})}-1 -\Re{\langle\rho|U|\rho\rangle})^2  }  \nonumber\\
&+ \frac{|\Im\braket{\rho|U|\rho} |^2}{8(1-\Re{\langle\rho|U|\rho\rangle}) + ( \frac{|1- \braket{\rho|U|\rho}|^2}{2(1 - \Im \braket{\rho|U|\rho})}-1 +\Re{\langle\rho|U|\rho\rangle})^2  } \Bigg].\label{eq:N_I_doublemeasurement_unitary}
\end{align}

The average value of $N_\Re$ and $N_\Im$ above over the unit circle computed with the error bounds is close to $1.2325\times N$, whereas if we assume $\ket{\rho}$ is an eigenstate, the result is $1.4805\times N$. The maximum value achieved is slightly above 2, $2.145\times N$. Statistical simulations show that the actual average over the unit circle is closer to $1.33 N_\Re$.

\subsection{Unitary state preparation}

In the circuit depicted in the main text,~\cref{fig:main_text_double_measurement}, the probability of measuring a 0 in the second register is
\begin{equation}
        P_\pm (0) =|\braket{\psi_\Re|\rho}|^2 = \frac{|1\pm \braket{\rho|U|\rho}|^2}{2(1 \pm \Re \braket{\rho|U|\rho})}.
\end{equation}
As a consequence,
    \begin{align}
        |1\pm \braket{\rho|U|\rho}|^2 = 2P_\pm (0)(1\pm\Re(\braket{\rho|U|\rho}))\\
        \Rightarrow|1\pm (\Re\braket{\rho|U|\rho}) + 1i\cdot (\Im\braket{\rho|U|\rho})|^2 = 2P_\pm (0)(1\pm\Re(\braket{\rho|U|\rho}))\\
        \Rightarrow |\Im\braket{\rho|U|\rho} |= \sqrt{2P_\pm (0)(1\pm\Re(\braket{\rho|U|\rho}))-|1\pm \Re\braket{\rho|U|\rho}|^2}
        \label{eq:abs(Im<U>)_unitary}
    \end{align}
where the $\pm$ depends on the result of the Hadamard test, $+$ for $\ket{0}$ and $-$ for $\ket{1}$. The sign of $\Re\braket{\rho|U|\rho}$ can only be determined in the Hadamard test. The error of this estimate is
\begin{align}
    \epsilon_{\Im, \pm}^2 &= \frac{4(1\pm\Re{\langle\rho|U|\rho\rangle})^2\epsilon_{S,\pm}^2 + [\pm 2P_{\pm}(0) \mp 2(1\pm\Re{\langle\rho|U|\rho\rangle})]^2\epsilon_H^2 }{4|\Im\braket{\rho|U|\rho} |^2}\\
    &= \frac{(1\pm\Re{\langle\rho|U|\rho\rangle})^2\epsilon_{S,\pm}^2 + [ P_{\pm}(0) - 1\mp\Re{\langle\rho|U|\rho\rangle})]^2\epsilon_H^2 }{|\Im\braket{\rho|U|\rho} |^2}
\end{align}
Substituting  $\epsilon_H^{2} \leq \frac{1}{2N}$, and $\epsilon_{S,\pm}^{2} \leq \frac{1}{2N_{\Re,\pm}} \approx \frac{2}{2N(1\pm\Re(\braket{\rho|U|\rho}))}$, we get
\begin{align}
    \epsilon_{\Im, \pm}^2 &= \frac{2(1\pm\Re{\langle\rho|U|\rho\rangle}) + [ \frac{|1\pm \braket{\rho|U|\rho}|^2}{2(1 \pm \Re \braket{\rho|U|\rho})} - 1\mp\Re{\langle\rho|U|\rho\rangle})]^2 }{|\Im\braket{\rho|U|\rho} |^2}\frac{1}{2N}.
\end{align}
The equivalent number of measurements would be
\begin{align}
N_{\Im} ={}& N \Bigg[ 1
+\frac{|\Im\braket{\rho|U|\rho} |^2}{2(1+\Re{\langle\rho|U|\rho\rangle}) + ( \frac{|1+ \braket{\rho|U|\rho}|^2}{2(1 + \Im \braket{\rho|U|\rho})}-1 -\Re{\langle\rho|U|\rho\rangle})^2  }  \nonumber\\
&+ \frac{|\Im\braket{\rho|U|\rho} |^2}{2(1-\Re{\langle\rho|U|\rho\rangle}) + ( \frac{|1- \braket{\rho|U|\rho}|^2}{2(1 - \Im \braket{\rho|U|\rho})}-1 +\Re{\langle\rho|U|\rho\rangle})^2  } \Bigg].\label{eq:N_I_doublemeasurement}
\end{align}
In this unitary case, we need $\times 1.49$ less circuits to obtain the same information as the Hadamard test alone. This becomes $\times 2.45$ if the initial state is an eigenvalue. The maximum multiplicative efficiency gain is $\times 3$.

\section{Composition of qDRIFT and product formulas~\label{app:qDRIFT_&_PFs}}
qDRIFT may be combined with product formulas, and used to simulate just some terms in the Hamiltonian CDF~\cite{hagan2023composite}. Since the cost functions are very different, this could help to simulate a large number of terms in the Hamiltonian with small norms. Let $\hat{H} = \sum_{p=1}^{N_p} H_p + \sum_q H_q$, where $H_q$ are the Hamiltonians to be simulated with qDRIFT. We shall consider the Lie Trotter product formula
\begin{align}\label{eq:composition_qdrift_product_formula}
    U_1(\tau) = \left(\prod_{i=1}^{N_p} e^{-iH_p\tau}\right)e^{-i\tau\sum_q H_q}
\end{align}
which may be used as the building block of more complex ones. The time evolution should be divided in $r = t/\tau$ segments, each of which uses calls to $U_1(\tau_j)$ $n_{U_1}$ times, for different $\tau_j$. $e^{-i\tau\sum_q H_q}$ shall be simulated with the qDRIFT technique. 
The simplest error analysis adds up the qDRIFT and $U_1(\tau)$ errors via a triangle inequality.

Deciding whether to allocate a term in the Hamiltonian to the product formula or the qDRIFT technique can be formulated in terms of a cost minimization problem subject to precision constraints. The qDRIFT error for the simulation of a single segment of length $\tau_j$ can be bounded as~\cite{campbell2019random}
\begin{align}
    \epsilon_{\tau_j} \leq \frac{2\lambda^2\tau^2_j}{r_{q,\tau_j}},
\end{align}
where $r_{q,\tau_j}$ is the number of terms exponentiated in that segment, and $\lambda$ the 1-norm of the terms implemented via qDRIFT. 
In each product formula segment, we may call $U_1(\tau)$ multiple times. Let us call $\tau_j$ the length of each of those calls.
Since we want $r \sum_j \epsilon_{\tau_j} = \frac{t}{\tau}\sum_j\epsilon_{\tau_j} \leq \epsilon$, we select $r_{q,\tau}$ such that
\begin{align}
    \frac{2\lambda^2 t}{\tau} \sum_j\frac{\tau_j^2}{r_{q,\tau_j}} \leq \epsilon.
\end{align}
If we pick $r_{q,\tau_j} = r_{q,\tau}\frac{|\tau_j|}{\sum_k |\tau_k|}$ for some $r_{q,\tau}$ corresponding to the whole segment, then
\begin{align}
    \frac{2\lambda^2 t}{\tau} \sum_j\frac{\tau_j^2}{r_{q,\tau_j}}  &= \frac{2\lambda^2 t}{\tau} \sum_k |\tau_k| \sum_j\frac{ \tau_j^2}{r_{q,\tau}|\tau_j|}= \frac{2\lambda^2 t}{r_{q,\tau} \tau}\left(\sum_j |\tau_j|\right)^2\leq \epsilon,
\end{align}
which means $r_{q,\tau_j}$ should be selected such that
\begin{align}
   r_{q,\tau}\geq  \frac{2\lambda^2 t}{\epsilon \tau}\left(\sum_j |\tau_j|\right)^2.
\end{align}
Note that while $\sum_j \tau_j = \tau$, some $\tau_j$ may be negative. Thus the use of the absolute value above. The total cost of qDRIFT is lower bounded by
\begin{align}
  C_{\text{qDRIFT}} = r \cdot r_{q,\tau}+
  &\geq  \frac{2\lambda^2 t^2}{\epsilon}\frac{\left(\sum_j |\tau_j|\right)^2}{\tau^2}= \frac{2\lambda^2 t^2}{\epsilon}\left(\sum_j |a_j|\right)^2,
\end{align}
where $a_j :=\tau_j/\tau$.

In contrast, estimating the error in a product formula is significantly more convoluted, see e.g.~\cite{childs2021theory}. The error per segment of length $\tau$ will be a key input to the cost. Once known, the cost of the product formula based on~\cref{eq:composition_qdrift_product_formula} is
\begin{align}
    C_{\text{pf}} = r\cdot n_{U_1}\cdot N_p = \frac{n_{U_1}N_p t}{\tau}.
\end{align} 
The cost of the full simulation is the sum of $C_{\text{pf}}$ and $ C_{\text{qDRIFT}}$.

\section{Accuracy of rotations}

In the literature there have been multiple proposals to implement the single qubit rotations that constitute the backbone of many Trotter algorithms, including the one one presented here. In the main text, we highlighted that restricting the accuracy of the rotations is equivalent to implementing exact Hamiltonian simulation of an effective Hamiltonian given by the approximated coefficients. This is true for unitary implementations of the rotations, including the Gidney adder~\cite{gidney2018halving} and the Ross and Selinger method~\cite{ross2014optimal}, and can be treated as perturbations to the Hamiltonian. However, more recent methods leverage quantum channels to reduce the necessary cost to achieve a certain precision. Perhaps the most performant such method is the mixed fallback method in Ref.~\cite{kivlichan2018quantum}. In this appendix we briefly analyze the effect of such method in the resulting spectrum.

The mixed fallback method synthesizes each rotation as a probabilistic combination between an overrotation and underrotation, each with its own fallback channel
\begin{equation}\label{eq:mixed_fallback}
    \mathcal{Z}_\theta (\rho) \approx pq_1\mathcal{Z}_{\theta + \delta_1}(\rho) + (1-p)q_2\mathcal{Z_{\theta + \delta_2}}(\rho) + p(1-q_1)\mathcal{B}_1(\rho) + (1-p)(1-q_2)\mathcal{B}_2(\rho).
\end{equation}
$p$ is chosen to be 
\begin{equation}
    p = \frac{q_2\sin(2\delta_2)}{q_2\sin(2\delta_2)-q_1\sin(2\delta_1)}
\end{equation}
for $\sin(\delta_1) \leq 0\leq \sin(\delta_2)$.
$\mathcal{B}_1$ and $\mathcal{B}_2$ represent fallback channels, implemented via unitary mixing with twirling. For each, we find two rotations, an over and an under rotation again, such that
\begin{align}
    \mathcal{B}_j(\rho)= p_j\mathcal{T}_{U_{j,1}}(\rho) + (1-p_j)\mathcal{T}_{U_{j,2}}(\rho),\\
    \mathcal{T}_{U_{j,k}}(\rho) = \frac{1}{4}\sum_{\sigma\in\{I, S, Z, S^\dagger\}} (\sigma U_{j,k} \sigma^\dagger)\rho(\sigma U_{j,k}^\dagger \sigma^\dagger).
\end{align}
Each twirl mixture yields a Pauli channel error (see Lemma 3.14 in Ref.~\cite{kliuchnikov2023shorter}). Unfortunately, the error in the mixed fallback method is not a Pauli channel error. However, it is possible to derive parameters $\delta_j$, $p_j$ and $q_j$ for $j\in\{1,2\}$ such that we will implement the desired Pauli Z rotation to a given diamond norm error $\epsilon_{\text{rot}}$. Our goal is to underestand how such $\epsilon_{\text{rot}}$ affects the time signal and spectrum beyond the perturbation of the Hamiltonian that we already discussed in the main text.

To model this effect, we first analyze how the error accumulates over the course of the simulation. The total evolution is a composition of a large number of rotations in discrete Trotter steps. The key assumption we make is that the incoherent error introduced at each step is independent of the errors in previous steps. This is a well-justified Markovian approximation, as the underlying gate synthesis errors are typically not correlated in time. This type of small, independent error accumulation leads to a gradual loss of phase coherence in the quantum state.

This loss of coherence can be rigorously connected to the diamond norm error
\begin{equation}
    \|\mathcal{R_Z}-\mathcal{\tilde{R}_Z}\|_\diamond \leq \epsilon.
\end{equation}
The diamond norm is an upper bound to the trace norm. The trace distance, on the other hand, bounds the infidelity of the quantum channel
\begin{equation}
    1-F \leq \frac{1}{2} \|\mathcal{R_Z}(\rho)-\mathcal{\tilde{R}_Z}(\rho)\|_1\leq \frac{1}{2} \|\mathcal{R_Z}(\rho)-\mathcal{\tilde{R}_Z}(\rho)\|_\diamond \leq \frac{\epsilon_{\text{rot}}}{2}.
\end{equation}
Thus, each rotation induces an infidelity upper bounded by $\epsilon_{\text{rot}}$. If one Trotter step needs $n_{\text{step}}$ single qubit rotations, the fidelity squared after one Trotter step might be bounded as
\begin{equation}
    F_{\text{step}}^2 = (1-\epsilon_{\text{rot}}/2)^{2 n_{\text{step}}}= e^{-n_{\text{step}}\epsilon_{\text{rot}}} + O(\epsilon_{\text{rot}}).
\end{equation}
Let us now evaluate how the infidelity affects the value of the time signal and spectrum. The time signal generated by the effective Hamiltonian in a Trotter formula is
\begin{equation}
   \tilde{G}(\tau j) = \Re(\braket{m_\rho|e^{-i \hat{H}'  \tau j}|m_\rho}) + i \Im(\braket{m_\rho|e^{-i \hat{H}'  \tau j}|m_\rho}).
\end{equation}
If we assume the error is Markovian and a combination of dephasing (due to different values of $\delta_1$ and $\delta_2$ in~\cref{eq:mixed_fallback}) and depolarization (due to the mixed unitary Pauli channels), then the time signal $\tilde{G}(\tau j)$ will be dampened by a factor $e^{-\Gamma \tau j}$ that captures the loss of fidelity described above. Let us denote by $\Xi_i^j$ with $i,j\in\{+,-\}$ the approximate density matrix entangled with $\ket{j}\bra{i}$ from the imperfect evolution in the Hadamard test. If we model the decoherence by a completely decohering channel, we would get
\begin{align}
    2P(0)-1 &= 2\text{Tr}\left((\ket{0}\bra{0}\otimes \bm{1})\ \left(\frac{1}{2}\ket{+}\bra{+}\otimes \ket{m_\rho} \bra{m_\rho} + \frac{1}{2}\ket{-}\bra{-}\otimes \Xi_{-}^{-} \right.\right.\nonumber\\
    &\quad \left.\left.+ \frac{1}{2}\ket{-}\bra{+}\otimes \Xi_{+}^{-} + \frac{1}{2}\ket{+}\bra{-}\otimes \Xi_{-}^{+}\right)\right)-1\\
    &\approx F_{\text{step}}^{2\tau j/\Delta}  \text{Tr}\left((\ket{0}\bra{0}\otimes \bm{1})\ \left(\ket{+}\bra{+}\otimes \ket{m_\rho} \bra{m_\rho} + \ket{-}\bra{-}\otimes e^{-i \hat{H}'  \tau j}\ket{m_\rho}\bra{m_\rho}e^{+i \hat{H}'  \tau j}\right.\right.\\
    &\left.\left.+\ket{+}\bra{-}\otimes \ket{m_\rho}\bra{m_\rho}e^{+i \hat{H}'  \tau j}+ \ket{-}\bra{+}\otimes e^{-i \hat{H}'  \tau j}\ket{m_\rho}\bra{m_\rho}\right)\right)\\
    &+ (1-F_{\text{step}}^{2\tau j/\Delta} ) \text{Tr}\left((\ket{0}\bra{0}\otimes \bm{1})\ \left(\ket{+}\bra{+}\otimes \ket{m_\rho} \bra{m_\rho} + \ket{-}\bra{-}\otimes \frac{\bm{1}}{d}\right)\right)-1\\
    &= F_{\text{step}}^{2\tau j/\Delta}  \text{Tr}\left((\ket{0}\bra{0}\otimes \bm{1})\ \left(\ket{+}\bra{-}\otimes \ket{m_\rho}\bra{m_\rho}e^{+i \hat{H}'  \tau j}+ \ket{-}\bra{+}\otimes e^{-i \hat{H}'  \tau j}\ket{m_\rho}\bra{m_\rho}\right)\right)\\
    &= F_{\text{step}}^{2\tau j/\Delta}  \Re(\braket{m_\rho|e^{-i\hat{H}'  \tau j}|m_\rho}),
\end{align}

and similarly for the imaginary component.
$\Gamma $ can then be upper bounded by $\frac{n_{\text{step}}\epsilon_{\text{rot}}}{\Delta}$. The expression of the spectrum at any given frequency is
\begin{equation}
     - \eta \operatorname{Im} G_\rho(\omega, H) = \frac{\eta \tau}{2\pi}\sum_{j=-\infty}^{+\infty} e^{-|j|\tau \eta}\tilde{G}(\tau j) e^{+ij\tau \omega}.
\end{equation}

This additional time-domain decay corresponds to convolving the spectrum with a Lorentzian function, adding to the intrinsic peak width $\eta$. The new effective broadening is $\eta_{\text{eff}}:= \eta + \Gamma \leq \eta + \frac{n_{\text{step}}\epsilon_{\text{rot}}}{\Delta}$. Thus, the effect of the incoherent imperfections in the accuracy of the rotation is to broaden the peaks in the spectrum. This leads to a critical trade-off when choosing the Trotter step size, $\Delta $, at fixed $\epsilon_{\text{rot}}$: decreasing $\Delta $ reduces the coherent Trotter error (improving peak positions), but it increases the decoherence rate $\Gamma$ because we need more rotations, causing more peak broadening.

If we want to make $\Gamma$ be a small fraction of $\eta$, we can derive the value of $\epsilon_{\text{rot}}$ that we need to target. For instance, let us take take $\eta = 0.05$ Ha. If we want $\Gamma\approx 0.1\eta$, and we select the $N=18$ spatial orbital MnLiO cluster together with the parameters in~\cref{tab:resources}, the resulting target precision is $\epsilon_{\text{rot}}\approx 9.8\times 10^{-8}$. 
The cost of each single-qubit rotation depends on the target precision. The expected cost of a $Z$ rotation using the mixed fallback method is~\cite{kliuchnikov2023shorter}
\begin{equation}
    C_{\text{rot}}(\epsilon_{\text{rot}}) = 0.53\log_2(\epsilon_{\text{rot}}^{-1}) + 4.86 \lesssim 18 \text{ T gates per rotation}.
\end{equation}
This is in contrast to the 10 Toffolis needed to approximate each rotation with the Gidney adder to $\epsilon_{\text{rot}} = 10^{-3}$~\cite{gidney2018halving}.

\end{document}